\documentclass{new_tlp}

\usepackage[utf8]{inputenc}
\usepackage[T1]{fontenc}
\usepackage[scaled=0.85]{beramono}
\usepackage[english]{babel}
\usepackage[final]{microtype}
\usepackage[autostyle]{csquotes}
\usepackage{amssymb}
\usepackage{amsmath}
\usepackage[usenames,dvipsnames]{xcolor}
\usepackage{wasysym} 
\usepackage[colorlinks=true,linkcolor=blue,citecolor=blue,urlcolor=blue,linktoc=all]{hyperref}
\hypersetup{final}
\usepackage{graphicx}
\graphicspath{{images/}}
\usepackage[misc]{ifsym}  
\usepackage[normalem]{ulem} 

\usepackage{multirow}

\usepackage[final]{listings}
\lstdefinelanguage{Maude}
{
morekeywords={protecting,pr,extending,ex,including,inc,op,ops,sort,sorts,subsort,subsorts,eq,ceq,rl,crl,mb,cmb,if,then,else,fi,mod,fmod,endm,endfm,is,ctor,constructor,comm,assoc,id:,var,vars,owise,and,or,sync,on,to,from,view,endv,th,endth,fth,endfth,in,aemod,endaem,emod,endem,ppt,ppts,ag,inh},
sensitive=true,
comment=[l]{---},
emph={\=},
alsoletter={-,:},
}
\def\mysize{\small}
\newcommand{\intt}[1]{\texttt{\mysize\upshape{#1}}}

\newcommand{\grt}{\mathrel{\triangleright}}
\newcommand{\pmap}{\mathrel{\ooalign{\hfil$\mapstochar\mkern5mu$\hfil\cr$\to$\cr}}}
\newcommand{\tdiam}{\mathop{\raisebox{0pt}{$\Diamond$}}}
\newcommand{\tbox}{\mathop{\raisebox{0pt}{$\Box$}}}
\newcommand{\tu}{\mathrel{\mathbf{U}}}
\newcommand{\tw}{\mathrel{\mathbf{W}}}
\newcommand{\tr}{\mathrel{\mathbf{R}}}
\newcommand{\tn}{\mathop{\Circle}}
\newcommand{\ltl}{\ensuremath{\mathrm{LTL}}}
\newcommand{\nonext}{\scalebox{-0.7}[.7]{$\varnothing$}}
\newcommand{\ltlmsp}{\mathrm{LTL}_{\nonext}(\Sigma,\Pi)}
\newcommand{\Land}{\bigwedge}

\newcommand{\limpl}{\mathrel{\rightarrow}}
\newcommand{\lequiv}{\mathrel{\leftrightarrow}}
\newcommand{\union}{\mathrel{\cup}}
\newcommand{\Union}{\bigcup}
\newcommand{\inters}{\mathrel{\cap}}

\newcommand{\aers}{\textnormal{\textup{\textsf{atEgRwSys}}}}
\newcommand{\ers}{\textnormal{\textup{\textsf{EgRwSys}}}}
\newcommand{\prs}{\textnormal{\textup{\textsf{RwSys}}}}
\newcommand{\aets}{\textnormal{\textup{\textsf{atEgTrStr}}}}
\newcommand{\ets}{\textnormal{\textup{\textsf{EgTrStr}}}}
\newcommand{\pts}{\textnormal{\textup{\textsf{TrStr}}}}
\newcommand{\calt}{\mathcal{T}}
\newcommand{\calr}{\mathcal{R}}
\newcommand{\calk}{\mathcal{K}}
\newcommand{\call}{\mathcal{L}}

\newcommand{\cals}{\mathcal{S}}
\newcommand{\cale}{\mathcal{E}}
\newcommand{\cala}{\mathcal{A}}

\newcommand{\calta}{{\calt\!,\cala}}
\newcommand{\compat}{\mathrel{|\hspace{-1.7mm}\approx}}
\newcommand{\sem}{\mathop{\mathrm{sem}}}
\newcommand{\simul}{\mathrel{\mathsf{S}}}

\newcommand{\rew}{\mathrel{\rightarrow}}
\newcommand{\splitop}{\mathop{\mathrm{split}}}

\newcommand{\rif}{\mathrel{\ \textsf{\small\upshape i\hspace{-0.7pt}f}\ }}
\newcommand{\trans}[1]{\mathrel{\ooalign{$-$\cr\hidewidth\hbox{$\big[\mkern-2mu$}\cr}\hbox{$\mkern3mu$}#1\hbox{$\mkern-3mu$}\mathrel{\ooalign{$\big]$\cr\hidewidth\hbox{$\rightarrow\mkern-13mu$}\cr}\hbox{$\mkern14mu$}}}}

\newtheorem{thm}{Theorem}
\newtheorem{prop}{Proposition}

\newtheorem{corol}{Corollary}
\newtheorem{defn}{Definition}

\usepackage{tikz}
\usetikzlibrary{arrows, arrows.meta, shapes.misc, decorations.pathmorphing, decorations.text}
\tikzset{>=stealth}
\tikzstyle{state} = [rounded rectangle, draw, fill=white, minimum size=2em]
\tikzstyle{trans} = [rectangle, draw, fill=gray!20, minimum size=1.9em]
\tikzset{sem/.style={-{Latex[length=2mm]}, decorate, decoration={snake, amplitude=.4mm, segment length=4mm, post length=1mm}, line width=0.3mm}}
\tikzset{split/.style={-{Latex[length=2mm]}, densely dashed, line width=0.3mm}}
\tikzset{sc/.style={-{Latex[length=2mm]}, double, line width=0.1mm}}
\tikzset{>=stealth}
\tikzstyle{box} = [rectangle, draw, minimum width=5em, minimum height=2em]
\tikzstyle{pnstate} = [rounded rectangle, draw, minimum size=1.5em]
\tikzstyle{pntrans} = [rectangle, draw, thick, fill=black, minimum width=6mm, inner ysep=2pt]

\title[Compositional verification in rewriting logic]
      {Compositional verification in rewriting logic%
      \thanks{Work partially supported by project S2018/TCS-4339 (BLOQUES-CM) co-funded by EIE Funds of the European Union and Comunidad de Madrid, and project PID2019-108528RB-C22 (ProCode-UCM) funded by Ministerio de Ciencia e Innovación (Spanish Government).}}

\author[Ó. Martín, A. Verdejo \and N. Martí-Oliet]
       {ÓSCAR MARTÍN, ALBERTO VERDEJO and NARCISO MARTÍ-OLIET
       \\ Facultad de Informática, Universidad Complutense de Madrid, Madrid, Spain
       \\ \email{\{omartins,jalberto,narciso\}@ucm.es}}

\begin{document}

\maketitle

\begin{abstract}
In previous work, summarized in this paper, we proposed an operation of parallel composition for rewriting-logic theories, allowing compositional specification of systems and reusability of components. The present paper focuses on compositional verification. We show how the assume/guarantee technique can be transposed to our setting, by giving appropriate definitions of satisfaction based on transition structures and path semantics. We also show that simulation and equational abstraction can be done componentwise. Appropriate concepts of fairness and deadlock for our composition operation are discussed, as they affect satisfaction of temporal formulas. We keep in parallel a distributed and a global view of composed systems. We show that these views are equivalent and interchangeable, which may help our intuition and also has practical uses as, for example, it allows global-style verification of a modularly specified system. \emph{Under consideration in Theory and Practice of Logic Programming (TPLP).}
\end{abstract}


\section{Introduction}

Rewriting logic~\cite{Mese92CondRwL} is a well established, logic-based formalism, useful, in particular, for the specification of concurrent and nondeterministic systems. There are ways, in this context, in which modularity can be achieved. The language Maude~\cite{ClavD+22Maude321}, for example, strongly based on rewriting logic, includes a powerful system of modules which promotes a good organization of the code. Besides, multicomponent or distributed systems are sometimes modeled as a multiset of objects and messages. However, a truly compositional specification was not possible. By that, we mean one in which each component is an independent rewrite system and composition is specified separately, allowing, for example, reusability of components. In previous work~\cite{MartVM20CompSpec}, we proposed an operation of parallel composition of rewrite systems to achieve precisely that. In the present paper, we show how a compositional specification written according to our proposal can be the object of compositional verification. Note that, in this work we often use \emph{rewrite system} as a shorthand for \emph{system specified using rewriting logic}.

The reasons for the convenience of a compositional approach to verification are well known: to avoid the state-explosion problem; because some systems are inherently compounds and it makes all sense to specify and verify them as such; because verified systems can be safely reused as library components.

There are two alternative views on the meaning of compositional specification, which lead to different needs for compositional verification. In one, a composed specification is seen as modeling a distributed system, of which probably only one component is under our control, and the aim of verification is to ensure that our component behaves appropriately in an appropriate environment. Global states are out of the question, and the behavior we focus on is that of our component. The assume/guarantee technique (see Section~\ref{sec:ag})) is designed to be helpful here.

In the other view, in contrast, the whole system is under our control, but working compositionally still makes sense for modular engineering. Then, the aim of compositional verification is to prove that each component behaves appropriately, not in a general, unknown environment, but in the particular one given by the rest of the components that we have also specified. The abstraction technique (see Section~\ref{sec:eq-abstr}) helps here: given a component and its environment, we can abstract either or both, and perform verification on the abstracted, simplified component and/or environment. Assume/guarantee may also help. For example, we use both techniques in the mutual exclusion example introduced in Section~\ref{sec:mutex}.

We consider special kinds of atomic and composed rewrite systems which we call \emph{egalitarian} and were introduced in our previous work~\cite{MartVM20CompSpec}. They are egalitarian in the sense that they give the same status to transitions and states. A composed egalitarian system is a set of independent but interacting atomic ones. We see them as modeling a distributed system. An egalitarian rewrite system, atomic or composed, can be translated into a standard rewrite system (called \emph{plain} in this work) by the operation that we call the \emph{split}. This allows to specify a system componentwise, translate the compound into a single plain system and, then, execute and verify the result monolithically using existing tools (the ones in Maude's toolset, for example). The relation between this monolithic verification and the compositional one using assume/guarantee is our Theorem~\ref{thm:ded-rule}.

We are interested in rewriting logic and, in this paper, in verifying systems specified using that logic. The underlying expectation is that a firm logical basis makes it easier to define, study, and implement modularity and composition. However, satisfaction of temporal formulas is defined on the transition structures which represent the semantics of the logical specifications. Thus, transition structures play a fundamental role in this paper, even if only as proxies for the main characters.

This is the plan of the paper. In Section~\ref{sec:example} we show and explain the compositional specification of three simple but illustrative examples. They are revisited later in the paper, but here they are meant as an informal introduction to our previous work on composition. Section~\ref{sec:background} contains a quick and mainly formal overview of our previous work. In Section~\ref{sec:paths}, we study execution paths, needed to define satisfaction of formulas. We consider paths in atomic components, sets of compatible paths from different components, and global paths, representing, respectively, local, distributed, and global behaviors. The contrast and equivalence between a local view and a global one is a constant throughout the paper. In Section~\ref{sec:ltl}, we describe the variant of the temporal logic LTL against which we verify our systems. In Section~\ref{sec:basic-sat}, we define basic satisfaction of temporal formulas based on paths, and show the relation between the distributed and the global views of satisfaction. In Section~\ref{sec:fair+deadlock}, we discuss the concepts of fairness and deadlocks, and their importance for compositional verification. In Section~\ref{sec:simul}, we consider the componentwise use of simulation and abstraction: simulating or abstracting a component induces the same on the whole system, with a potentially reduced effort. In addition, the abstracted system may be easier to verify. In Section~\ref{sec:ag}, we consider the assume/guarantee technique, which allows the verification of isolated individual components, ensuring thus that the result holds for whatever appropriate environment the component is placed in, and we show how it can be adapted to our setting. In Section~\ref{sec:add-ex}, we briefly present two additional examples of compositional specification and verification (fully discussed in~\cite{Mart21Thesis}) which are more complex and realistic than the toy ones used throughout this paper. Finally, Section~\ref{sec:closing} discusses related and future work and contains some closing remarks.

These are the points we think may be of special interest in this paper:
\begin{itemize}
\item
   We show how to work compositionally in rewriting logic, expanding and strengthening our previous work.
\item
   We keep in parallel, all throughout the paper, the distributed and the global (monolithic) views of satisfaction and related concepts, and show the equivalence of both views. We claim that keeping both views is worth the effort, both for our intuition and in practice.
\item
   We show how simulation and abstraction can be performed compositionally.
\item
   We show that the assume/guarantee technique can be transposed to our setting.
\item
   Our definition of assume/guarantee satisfaction (inductive, but not relying on the \emph{next} temporal operator) is new, to the best of our knowledge.
\item
   We give path-based definitions of deadlock and fairness, discuss how they impact the verification tasks, and show how to deal with them in our setting.
\end{itemize}

\section{Examples}
\label{sec:example}

We introduce here three examples of compositional specification. They are meant as a quick introduction to Maude and to our previous work~\cite{MartVM20CompSpec}, especially to compositional specification with the extended syntax we proposed. The formal definitions and results are in Section~\ref{sec:background}. Also, these examples set the base on which we later illustrate the techniques for compositional verification and simulation. They have been chosen to be illustrative, so they are quite simple. We are using Maude because of its availability, its toolset, and its efficient implementation. All the concepts and examples, however, are valid for rewriting logic in general.

The first example presents three buffers assembled in line. The second shows how to exert an external mutual-exclusion control on two systems, provided they inform on when they are visiting their critical sections. Later, this gives us the opportunity of using componentwise simulation and a very simple case of assume/guarantee. The third example concerns the well-known puzzle of a farmer and three belongings crossing a river. We compose a system implementing the mere rules of the puzzle with several other components implementing, in particular, two guidelines which prove to be enough to reach a solution. The assume/guarantee technique is later used on this system.

The complete specification for all the examples in this paper is available online~\cite{Mart20Web}. Our prototype implementation, able to deal with these examples, is also available there, though the reader is warned that, in its current state, it is not a polished tool but, rather, a proof of concept.

\subsection{Chained buffers}
\label{sec:buffers}

We model a chain of three buffers. We describe the system top-down. This is the specification of the composed system:
\begin{lstlisting}
sync BUFFER1 || BUFFER2 || BUFFER3
   on BUFFER1$isSending = BUFFER2$isReceiving
   /\ BUFFER2$isSending = BUFFER3$isReceiving .
\end{lstlisting}
The \lstinline!sync...on...! sentence is not standard Maude, but part of our extension. That sentence expects three Maude modules to exist, called \lstinline!BUFFER1!, \lstinline!BUFFER2!, and \lstinline!BUFFER3!, each defining the values of the so-called \emph{properties} mentioned in the \lstinline!on! part of the sentence: \lstinline!isSending! and \lstinline!isReceiving!. We use the \lstinline!$! sign to access a property defined in a Maude module. In words, that models a composed system in which the three buffers synchronize so that when one sends the next receives. The properties are assumed to be Boolean in this example, modeling the passing of tokens. Synchronizing on more complex values is also possible, as shown in other examples.

We call the result of the composition above \lstinline!3BUFFERS!. For this to be a complete model, we need to provide the specification of the internal workings of the three buffers, including the definition of the properties. There is no reason for the three buffers to be specified exactly the same. In principle, they even could be coded in different languages, as long as there is a way to access the values of the properties defined inside each of them. For the sake of simplicity, in this example the three modules are identical. This is the very simple code for each of them:
\begin{lstlisting}
sort State Trans .
ops idle gotToken : -> State .
ops receiving sending : -> Trans .
rl idle      =[ receiving ]=>  gotToken .
rl gotToken  =[ sending   ]=>  idle .
\end{lstlisting}
There are two states, represented by the \lstinline!State! constants \lstinline!idle! and \lstinline!gotToken!, and two transitions between them, represented by the \lstinline!Trans! constants \lstinline!receiving! and \lstinline!sending!. The keyword \lstinline!ops! introduces the declaration of operators with their arities. The singular \lstinline!op! can be used when only one operator is being declared. In this code, we are declaring constants, so the argument sorts are absent. The keyword \lstinline!rl! introduces each rewrite rule and the symbols \lstinline!=[! and \lstinline!]=>! separate the terms. We assume throughout the paper that the sort representing the states of the system is called \lstinline!State! and the one representing transitions is called \lstinline!Trans!. Also, it is convenient to have a supersort of both (not shown above), which we call \lstinline!Stage!. Usually, we omit declarations of sorts and operators when they are clear from context.

Readers knowledgeable of rewriting logic and Maude would expect the rules above to be written instead as:
\begin{lstlisting}
rl [receiving] : idle     => gotToken .
rl [sending]   : gotToken => idle .
\end{lstlisting}
Here, \lstinline!receiving! and \lstinline!sending! are rule labels. The syntax we use does not only consists of moving the label to the middle of the rule. In our case, \lstinline!receiving! and \lstinline!sending! are not labels, but algebraic terms of sort \lstinline!Trans!, in the same way that \lstinline!idle! and \lstinline!gotToken! are terms of sort \lstinline!State!. In general, both \lstinline!State!s and \lstinline!Trans!s can be terms of any algebraic complexity. Other examples below make this clearer.

We call these rules \emph{egalitarian}, because transitions are represented by terms, the same as states. The rewrite systems which include them are also called \emph{egalitarian}. More precisely, each buffer is an \emph{atomic egalitarian rewrite system}. The result of their composition is still called \emph{egalitarian}, but not atomic.

As illustrated above, the way we have chosen to specify composition of systems is by equality of \emph{properties}. These are functions which take values at each state and transition of each component system. The properties of a system provide a layer of isolation between the internals of each component and the specification of the composition. This is similar to the concept of ports in other settings. It is important that properties are defined not only on states, but also on transitions, because synchronization is more often than not specified on them. That is why we have developed egalitarian systems in which transitions are promoted to first-class citizenship. 

We declare and define two properties in each buffer:
\begin{lstlisting}
ppt isReceiving isSending : -> Bool .
eq isReceiving @ receiving = true .
eq isReceiving @ G = false [owise] .
eq isSending @ sending = true .
eq isSending @ G = false [owise] .
\end{lstlisting}
The sentence introduced by the keyword \lstinline!ppt! is part of our extended syntax, as is the symbol \lstinline!@! representing the evaluation of a property on a state or transition. Thus, these lines declare two Boolean properties and define by means of equations (introduced by the keyword \lstinline!eq!) their values at each state and transition. The fact that \lstinline!receiving! and \lstinline!sending! are algebraic terms allows their use in equations.

The attribute \lstinline!owise! (short for \emph{otherwise}) in two of the equations is an extralogical feature of Maude: that equation is used whenever the term being reduced matches the left-hand side and the case is not dealt with by other equations. The variable \lstinline!G!, whose declaration is not shown, has sort \lstinline!Stage!, so that all properties evaluate to \lstinline!false! except in the two cases explicitly set to \lstinline!true!.

Any property defined in a component can be used as well as a property for the resulting composed system. In this case, the properties \lstinline!isReceiving! in \lstinline!BUFFER1! and \lstinline!isSending! in \lstinline!BUFFER3! are defined but not used for synchronization. Those properties can be useful if the composed module \lstinline!3BUFFERS! is used in turn as a component to be synchronized with other modules.

It is a common case that a property is defined to be true exactly at one state or transition and false everywhere else, as above. This calls for some syntactic shortcut to help the user. We do not discuss in this paper how to implement such shortcuts (of which this is not at all the only possible one), and our prototype implementation does not include them.

The execution of the composed system \lstinline!3BUFFERS! consists in the independent execution of each of its three components, restricted by the need to keep the equality between properties. To that composed system, the operation we call the \emph{split} can be applied to obtain an equivalent standard rewrite system. The resulting split system has as states triples like \lstinline!< idle, gotToken, idle >!, formed from the states of the components, and has rewrite rules like
\begin{lstlisting}
rl < idle, gotToken, idle > => < idle, sending, receiving > .
\end{lstlisting}
The split is named after this translation of each rule into two \emph{halves}. The term \emph{split} is also used later to describe related translations, though in some of those cases there is nothing split in the literal sense. The split is formally defined in Section~\ref{sec:split}. We usually do not care to show the internal appearance of a split system, but are only interested in the fact that it represents in a single system the global behavior of the composition.

\subsection{Mutual exclusion}
\label{sec:mutex}

Consider a very simple model of a train, which goes round a closed railway in which there are three stations and a crossing with another railway. We use the three stations as the states of our model, and there are three transitions for moving between them. Using our extended syntax, we model it with the rule:
\begin{lstlisting}
crl atStation N   =[ comingFrom N ]=>   atStation (N + 1)   if N < 2 .
\end{lstlisting}
The keyword \lstinline!crl! introduces a conditional rewrite rule. We omit the needed declarations for the integer variable \lstinline!N! and the constructors \lstinline!atStation! and \lstinline!comingFrom!.

The stations are numbered 0 to 2. But the transit from station 2 to 0 is different, because it passes through the crossing:
\begin{lstlisting}
rl atStation 2   =[ crossing ]=>   atStation 0 .
\end{lstlisting}

Indeed, we have two trains, modeled in this example by the same specification, but as two separate components. They share the crossing, so we need safety in the access to it. To this aim, we define for each train a Boolean property \lstinline!isCrossing! to be true at the transition \lstinline!crossing! and false everywhere else:
\begin{lstlisting}
ppt isCrossing : -> Bool .
eq isCrossing @ crossing = true .
eq isCrossing @ G = false [owise] .
\end{lstlisting}
We call the two systems thus defined \lstinline!TRAIN1! and \lstinline!TRAIN2!.

The mutex controller for safe access to the crossing is specified by these two rules:
\begin{lstlisting}
rl idle   =[ grants 1 ]=>   idle .
rl idle   =[ grants 2 ]=>   idle .
\end{lstlisting}
We call this system \lstinline!MUTEX! and define in it the parametric Boolean property \lstinline!isGranting!, which is defined to be true at the respective transitions and false everywhere else:
\begin{lstlisting}
ppt isGranting : Nat -> Bool .
eq isGranting(I) @ (grants I) = true .
eq isGranting(I) @ G = false [owise] .
\end{lstlisting}

The final system is the composition of the two trains and \lstinline!MUTEX! so that each \lstinline!isCrossing! property is synchronized with the corresponding \lstinline!isGranting! one:
\begin{lstlisting}
sync TRAIN1 || TRAIN2 || MUTEX
   on TRAIN1$isCrossing = MUTEX$isGranting(1)
   /\ TRAIN2$isCrossing = MUTEX$isGranting(2) .
\end{lstlisting}

In due time, in Sections~\ref{sec:mutex-cont} and~\ref{sec:mutex-cont2}, we will show how we can use simulation to work with even simpler models of the trains, and how we can justify that mutual exclusion holds for the composed system.

We want to insist in the value of modularity in our examples. The system \lstinline!MUTEX! with its two properties can be used unchanged to control any two given systems, as long as they inform, by means of properties, of their being in their critical section. For general systems, the synchronization instruction would look something like
\begin{lstlisting}
sync ONE-SYSTEM || ANOTHER-SYSTEM || MUTEX
   on ONE-SYSTEM$isInCS = MUTEX$isGranting(1)
   /\ ANOTHER-SYSTEM$isInCS = MUTEX$isGranting(2) .
\end{lstlisting}
Mutual exclusion between the two systems, whatever they are, is guaranteed by \lstinline!MUTEX! satisfying the appropriate formula---see Section~\ref{sec:mutex-cont2}.

We find cases like this of particular interest. We mean a component controlling others and imposing its behavior (mutual exclusion in this case) on the compound. This is the idea behind strategies, controllers, coordination, etc. In contrast, in the example of the chained buffers in Section~\ref{sec:buffers}, the composed behavior is emergent. Our next example involves both techniques.

\subsection{Crossing the river}
\label{sec:river}

For a quick reminder, this is the statement of the puzzle. A farmer has got a wolf, a goat and a cabbage, and needs to cross a river using a boat with capacity for the farmer and, at most, one of the belongings. The wolf and the goat should not be left alone, because the wolf would eat the goat. In the same way, the goat would eat the cabbage if left unattended. The goal is to get the farmer and the three belongings at the opposite side of the river safely.

Our specification consists of two rules: one encompasses all possible ways the farmer can cross the river; the other represents eating. This is the rule for a crossing, explained below:
\begin{lstlisting}
rl   farmer B? II1  |~|     II2
  =[ II1            | B? >  II2           ]=>
     II1            |~|     farmer B? II2 .
\end{lstlisting}
Each state term contains the symbol \lstinline!|~|! representing the river. To each side of this symbol there is a set of items, which may include the farmer and the three belongings, respectively represented by the constants \lstinline!farmer!, \lstinline!wolf!, \lstinline!goat!, and \lstinline!cabbage!. Also, there is always a special item \lstinline!mark! which marks the side that the farmer is trying to reach with her belongings. Thus, the initial state is defined like this:
\begin{lstlisting}
eq init = farmer wolf goat cabbage |~| mark .
\end{lstlisting}

The variables \lstinline!II1! and \lstinline!II2! are sets of items which, in particular, may be empty. The sort of the variable \lstinline!B?! is \lstinline!MaybeBelong!, that is, either one of the three belongings or the special value \lstinline!noBelong!. Indeed, \lstinline!noBelong! is also the identity element for sets of items. In this way, the transition term \lstinline!II1 | B? > II2! represents all possible crossings, with $\intt{B?}=\intt{noBelong}$ interpreted as the farmer crossing alone. The symbol \lstinline!|~|! is formally a commutative operator, so that the same rule represents movements from any side to the other. That rule is rather terse. Alternative specifications, using more than one rule, would probably be easier to grasp. That is not important for the main purpose of this paper, which has to do with composition.

The rule for eating is this one:
\begin{lstlisting}
rl   goat B II1       |~|  farmer II2
  =[ eating B II1     |~|  farmer II2 ]=>
     survivor(B) II1  |~|  farmer II2 .
\end{lstlisting}
Thus, when the goat and some other belonging are at one side with the farmer at the other side, eating can take place. The function \lstinline!survivor! is defined by these equations:
\begin{lstlisting}
eq survivor(cabbage) = goat .
eq survivor(wolf) = wolf .
\end{lstlisting}
Thus, the goat survives if the other belonging is the cabbage, but it dies (disappears from the state term) if the other belonging is the wolf.

Our specification does not require that eating happens as soon as it is possible, but only that it \emph{can} happen. So our aim is to avoid all danger and ensure a safe transit.

This was the specification of the rules of the game. We propose now two guidelines for the farmer to follow. The first is to avoid all movements which lead to a dangerous situation, that is, one with the goat and some other belonging left by themselves. The second is to avoid undoing the most recent crossing: for example, after crossing one way with the goat, avoid going back the other way with the goat again. These are both quite obvious guidelines to follow, and we hypothesize that they are enough to ensure that the farmer reaches the goal. As it turns out, the hypothesis is false, and we will need to strengthen the second guideline; but let us work with this for the time being.

The guidelines are enforced by avoiding certain transitions to be triggered. For that, we need to identify said transitions. First, the dangerous ones:
\begin{lstlisting}
ppt danger : -> Bool .
eq danger @ (goat B II1 | B? > II2) = true .
eq danger @ G = false [owise] .
\end{lstlisting}
The variable \lstinline!B! represents a belonging, while \lstinline!B?!, as before, can be either a belonging or \lstinline!noBelong!. In words: there is danger if the farmer is in the boat and the goat has been left alone with some other belonging.

We need to restrict the execution of \lstinline!RIVER! so that \lstinline!RIVER$danger = false! at all times. This is another instance where a syntactic shortcut would help, but also this requirement can be enforced by a composition with an appropriate controller.

Let us call the following system \lstinline!AVOID!. It is as simple as a system can possibly be:
\begin{lstlisting}
op init : -> State .
ppt avoid : -> Bool .
eq avoid @ init = false .
\end{lstlisting}
There is a single state, called \lstinline!init!, no transitions and no rules, and the property \lstinline!avoid! is always false. Thus, the composed system
\begin{lstlisting}
sync RIVER || AVOID
   on RIVER$danger = AVOID$avoid .
\end{lstlisting}
indeed avoids all situations at which \lstinline!danger! is true.

Implementing the other guideline, avoidance of the undoing of movements, requires one more step, because we need to, somehow, store the previous movement so as to be able to compare it with the potential new one. We are after a composed system like this
\begin{lstlisting}
sync RIVER || PREVIOUS
   on PREVIOUS$move = RIVER$move .
\end{lstlisting}
where \lstinline!RIVER! \emph{informs} the new system \lstinline!PREVIOUS! about the moves being made, and \lstinline!PREVIOUS! stores at each moment the latest move. We name this composed system \lstinline!RIVER-W-PREV!.

The new component \lstinline!PREVIOUS! needs only this rule:
\begin{lstlisting}
rl B?   =[ B? > B?' ]=>   B?' .
\end{lstlisting}
Its state sort is \lstinline!MaybeBelong!, that is, either actually one of the three belongings or the value \lstinline!noBelong!. In this case they are representing movements: the farmer crossing either with the specified belonging or alone. The transition term includes two such movements: the previous one and the new one. In this way, we can check them for equality when needed. To synchronize with the main system \lstinline!RIVER!, we use this property in \lstinline!PREVIOUS!:
\begin{lstlisting}
ppt move : -> MaybeMove .
eq move @ (B? > B?') = B?' .
eq move @ B? = noMove .
\end{lstlisting}
Correspondingly, we need this property in \lstinline!RIVER!:
\begin{lstlisting}
ppt move : -> MaybeMove .
eq move @ (II1 | B? > II2) = B? .
eq move @ G = noMove [owise] .
\end{lstlisting}
We need to include the new constant \lstinline!noMove! for when, indeed, no move is taking place.

Storing information about the past execution of the system is called \emph{instrumentation} and is a common technique in system analysis. This is another instance calling for syntactic sugar. As shown with the \lstinline!RIVER || PREVIOUS! example, it can be achieved by composition of atomic rewrite systems.

Whenever \lstinline!RIVER! is executing a crossing, \lstinline!PREVIOUS! is showing, in its transition term, the previous and the current moves, giving us the possibility of checking if they are equal:
\begin{lstlisting}
ppt undoing : -> Bool .
eq undoing @ (B? > B?) = true .
eq undoing @ G = false [owise] .
\end{lstlisting}

Now, we need to restrict \lstinline!RIVER-W-PREV! so as to avoid undoing movements. For that, we can use \lstinline!AVOID!, as above. But we need two instances of that system, one to avoid danger, the other to avoid undoings, to which we refer as \lstinline!AVOID1! and \lstinline!AVOID2!.

At the end, the system we are interested in is
\begin{lstlisting}
sync RIVER-W-PREV || AVOID1 || AVOID2
   on RIVER$danger = AVOID1$avoid
   /\ PREVIOUS$undoing = AVOID2$avoid .
\end{lstlisting}

This completes the specification of the system. Later in the paper, in Section~\ref{sec:river-cont}, we show how to verify that it leads to a solution\dots\ or, rather, that it does not. But we will also show a sufficient strengthening of the concept of undoing.

As in the previous examples, we want to draw the reader's attention to the modularity of our specification. Some previous treatments of this problem in rewriting logic~\cite{PaloMV05PlayMaude,RubiM+21StratCTL} used several rules to model the different ways of crossing. But this is irrelevant to us, because any specification that defines the properties \lstinline!move! and \lstinline!danger! will do as well.

\section{Background}
\label{sec:background}

This section is a formal summary of our previous work on the synchronous composition of rewrite systems~\cite{MartVM20CompSpec}. Detailed explanations and proofs can be found there. This whole section is quite theoretical, consisting of many definitions and a few propositions, to complement the informal and example-based introduction in Section~\ref{sec:example}.

We define below a number of structures and systems. This is a list of them with the abbreviations we use to refer to them:
\begin{tabbing}
\hspace{1mm} \= \aers: \hspace{1mm} \= atomic egalitarian rewrite systems \\
\> \ers:  \> egalitarian rewrite systems \\
\> \prs:  \> plain rewrite systems \\
\> \aets: \> atomic egalitarian transition structures \\
\> \ets:  \> egalitarian transition structures \\
\> \pts:  \> plain transition structures
\end{tabbing}

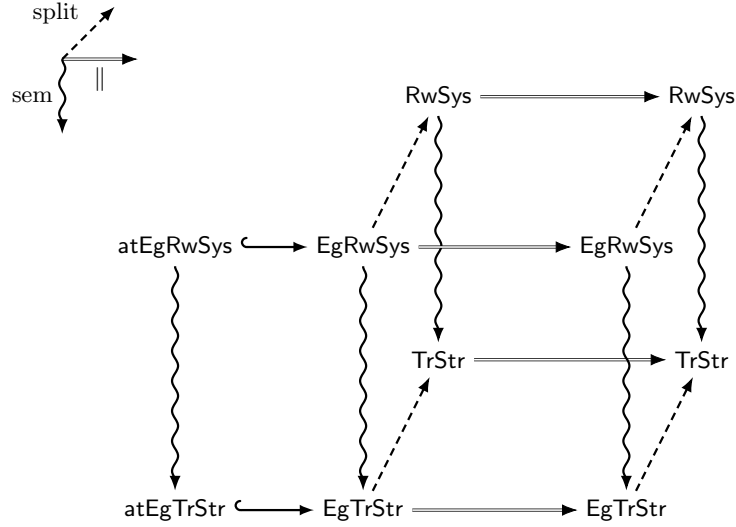
\begin{figure}[t]
\centering
\begin{tikzpicture}[auto, every text node part/.style={align=left}]
\node [rectangle] (atEgTrStr) at (0.5, 0) {\aets};
\node [rectangle] (EgTrStr) at (3, 0) {\ets};
\node [rectangle] (EgTrStr') at (6.5, 0) {\ets};
\node [rectangle] (TrStr) at (4, 2) {\pts};
\node [rectangle] (TrStr') at (7.5, 2) {\pts};

\node [rectangle] (atEgRwSys) at (0.5, 3.5) {\aers};
\node [rectangle] (EgRwSys) at (3, 3.5) {\ers};
\node [rectangle] (EgRwSys') at (6.5, 3.5) {\ers};
\node [rectangle] (RwSys) at (4, 5.5) {\prs};
\node [rectangle] (RwSys') at (7.5, 5.5) {\prs};

\draw (EgRwSys) edge [split] (RwSys);
\draw (EgRwSys') edge [split] (RwSys');

\draw (EgTrStr) edge [split] (TrStr);
\draw (EgTrStr') edge [split] (TrStr');

\draw (atEgRwSys) edge [sem] (atEgTrStr);
\draw (EgRwSys) edge [sem] (EgTrStr);
\draw (RwSys) edge [sem] (TrStr);
\draw (RwSys') edge [sem] (TrStr');

\draw (TrStr) edge [sc] (TrStr');
\draw (EgTrStr) edge [sc] (EgTrStr');
\draw (RwSys) edge [sc] (RwSys');
\draw (EgRwSys) edge [sc] (EgRwSys');

\draw (EgRwSys') edge [sem] (EgTrStr');

\draw (-1, 6) edge [sc] node [below] {$\|$} (0, 6);
\draw (-1, 6) edge [sem] node [left] {$\sem$} (-1, 5);
\draw (-1, 6) edge [split] node {split} (-0.3, 6.7);

\draw [right hook-latex, line width=0.3mm] (atEgRwSys) edge (EgRwSys);
\draw [right hook-latex, line width=0.3mm] (atEgTrStr) edge (EgTrStr);
\end{tikzpicture}
\caption{The types of systems we use and their relations.}
\label{fig:polyh}
\end{figure}

The polyhedron in Figure~\ref{fig:polyh} shows the whole set of structures and systems with their related maps.
Slanted dashed arrows represent the several concepts of split, that is, of obtaining plain transition structures or rewrite systems from egalitarian ones. Double horizontal arrows represent synchronous composition of systems or structures: composing systems or structures of the same kind produces another one of the same kind. Downward snake arrows represent semantic maps, assigning transition structures to rewrite systems. The two horizontal hooked arrows on the left represent inclusion: atomic systems and structures are particular cases of general systems and structures, respectively. All the elements in the diagram are defined below, and better explained in our previous paper~\cite{MartVM20CompSpec}.

\subsection{Egalitarian structures and systems}

As we mentioned above, we use transition structures (of particular types) as semantics for our rewrite systems. In due time, we define execution paths for transition structures, and satisfaction based on those paths. In this section we define atomic egalitarian transition structures, atomic egalitarian rewrite systems, the semantic relation between them, and their compositions.

\begin{defn}[atomic egalitarian transition structure]
\label{def:aets}
An \emph{atomic egalitarian transition structure} is a tuple $\calt=(Q,T,\rew,P,g_0)$, where:
\begin{itemize}
\item
   $Q$ is the set of states;
\item
   $T$ is the set of transitions;
\item
   ${\rew} \subseteq (Q\times T) \union (T\times Q)$ is the bipartite adjacency relation;
\item
   $P$ is the set of properties, each one a total function $p$ from $Q\union T$ to some codomain $C_p$;
\item
   $g_0 \in Q\union T$ is the initial state or transition.
\end{itemize}
We refer to the elements of $Q\union T$ as \emph{stages}.
The class of atomic egalitarian transition structures is denoted by \aets.
\end{defn}

The adjacency relation allows for several arrows in and out of a transition, as well as a state. The egalitarian goal also mandates that not only an initial state is possible, but also an initial transition. We use variables typically called $g$, with or without ornaments, to range over stages.

The definition of an atomic egalitarian transition structure is almost identical to that of a Petri net. The difference, however, is in the semantics: we are interested in a simple path semantics, instead of sets of marked places. This is better explained in Section~\ref{sec:paths}.

In the definitions below, for a given signature $\Sigma$, we denote by $T_\Sigma$ the set of terms on $\Sigma$, by $T_\Sigma(X)$ the terms with sorted variables from the set $X$, and by $T_{\Sigma,s}$ and $T_\Sigma(X)_s$ the terms of sort $s$ from the respective sets. Finally, $\Sigma|_s=\{f:s\to s' \mid \text{ for some } s'\in S\}$ denotes the set of totally defined unary operators in $\Sigma$ with domain $s\in S$.

\begin{defn}[atomic egalitarian rewrite system]
\label{def:aers}
An \emph{atomic egalitarian rewrite system} is a tuple $\calr=(S,{\le},\Sigma,E,R)$, where:
\begin{itemize}
\item
   $(S,{\le})$ is a poset of sorts. We assume $\intt{State}, \intt{Trans},\intt{Stage}\in S$ with $\intt{State}\le\intt{Stage}$ and $\intt{Trans}\le\intt{Stage}$. The terms of sort \lstinline!Stage! are called \emph{stages}.
\item
   $\Sigma$ is a signature of operators (and constants) $f : \omega \to s$ for some $\omega\in S^*$ and $s\in S$. We assume there is a constant $\intt{init}\in\Sigma$ of sort \lstinline!Stage!.
\item
   $E$ is a set of left-to-right oriented equations
   \[t=t' \rif C\]
   where $t,t'\in T_\Sigma(X)_s$ for some $s\in S$ and the condition $C$ (which may be absent) is a conjunction $\Land_i u_i=u_i'$ of equational conditions, for $u_i,u_i'\in T_\Sigma(X)_{s_i}$ for some $s_i\in S$.
\item
   The set $R$ contains egalitarian rules, that is, rules of the form
   \[u \trans{t} u' \rif C\]
   where $u,u'\in T_\Sigma(X)_\intt{State}$, $t\in T_\Sigma(X)_\intt{Trans}$ and $C$ (which may be absent) is as above.
\end{itemize}
We also refer as \emph{signature} to the triple $(S,{\le},\Sigma)$. A \emph{property} is any element of $\Sigma|_\intt{Stage}$, that is, any unary operator in $\Sigma$ totally defined on \lstinline!Stage! terms.
\end{defn}

The main point in which we depart from the standard definitions of rewrite system (often called rather \emph{rewrite theory})~\cite{Mese92CondRwL} is that our rules are egalitarian, by which we mean that they include an explicit transition term. Properties are also a non-standard ingredient. As a passing note, we have shown~\cite[Section~6.2.4]{Mart21Thesis} that requiring properties to be totally defined, as we do, is not a meaningful restriction.

In Maude, and in our examples in this paper, equations are introduced by the keywords \lstinline!eq! or \lstinline!ceq!, and rules by \lstinline!rl! or \lstinline!crl!; in each case the {\ttfamily\bfseries\color{red!50!black}c} form is used when conditions are present. The signature is represented by sentences with keywords \lstinline!sort!, \lstinline!subsort!, and \lstinline!op! or \lstinline!ops!, though we often omit such sentences in the examples in this paper.

Some of the definitions and results that follow are very similar for transition structures and for rewrite systems. In particular, the synchronization mechanism is the same for one and the other. To minimize repetition, we deal with both of them jointly as much as possible. We refer to them in abstract as \emph{systems} and with the letter $\cals$.

We compose atomic systems and structures to create complex ones. In all this paper we consider each system to be its own namespace, so that the sets of properties, sorts and operators from different systems are disjoint.

\begin{defn}[suitable synchronization criteria]
\label{def:suitable}
Given a set of atomic structures or systems, one for each $n=1,\dots,N$, either all of them in \aets\ or all of them in \aers, each with set of properties $P_n$, a set of \emph{synchronization criteria} for them is a set $Y\subseteq\Union_nP_n\times\Union_nP_n$.

We say that a set $Y$ of synchronization criteria is \emph{suitable} if it satisfies the following conditions. For transition structures, we require that, if $(p,p')\in Y\inters (P_m\times P_n)$, for some $m,n\in\{1,\dots,N\}$, with $p : Q_m\union T_m\to C$ and $p' : Q_n\union T_n\to C'$, then the elements in $C$ and $C'$ can be compared for equality. Correspondingly, for rewrite systems $\calr_n=(S_n,\le_n,\Sigma_n,E_n,R_n)$, we require that, if $(p,p')\in Y\inters(P_m\times P_n)$, with $p : \intt{Stage}_m \to s$ and $p' : \intt{Stage}_n \to s'$, then there exists a sort $s_0$, common to $\calr_m$ and $\calr_n$, with $s_m\le_m s_0$ and $s_n\le_n s_0$, and an equational theory $\cale_0$ of $s_0$, included as subtheory in both $\calr_m$ and $\calr_n$, in which the values of $p$ and $p'$ can be checked for equality.
\end{defn}

To be precise, we should require that $\cale_0$ be \emph{embedded} (rather than \emph{included}) by means of injective maps into the equational theories of $\calr_m$ and $\calr_n$. In that way, the namespaces of different systems are kept disjoint. While it is technically imprecise, we use the shorthand of saying that $\cale_0$ is the \emph{common equational theory} of $s_0$.

\begin{defn}[synchronous composition]
\label{def:sync-comp}
The \emph{synchronous composition} of $\cals_n$ for $n=1,\dots,N$,  either all of them in \aets\ or all of them in \aers, with respect to the suitable synchronization criteria $Y$ is denoted by $\|_Y\{\cals_n\mid n=1,\dots,N\}$, or usually just $\|_Y\cals_n$. From now on, whenever we write $\|_Y\cals_n$, we are assuming $Y$ is suitable. When only two components are involved, we usually write $\cals_1\|_Y\cals_2$.
\end{defn}

\begin{defn}[egalitarian structures and systems]
We define the classes of \emph{egalitarian transition structures}, denoted by \ets, and, respectively, of \emph{egalitarian rewrite systems}, denoted by \ers, as the smallest ones that contain \aets\ or, respectively, \aers, and are closed with respect to the synchronous composition operation described above.
\end{defn}

We need to consider a notion of equivalence: the one given by the different ways of composing the same components. For example, $(\cals_1\|_{Y}\cals_2)\|_{Y'}\cals_3$ is equivalent to $\|_{Y\union Y'}\{\cals_1, \cals_2, \cals_3\}$.

\begin{defn}[equivalent structures and systems]
\label{def:equiv-sys}
The \emph{set of atomic components} of an egalitarian transition structure or rewrite system is:
\begin{itemize}
\item $\mathop{\textrm{atoms}}(\cals) = \{\cals\}$ \; if $\cals$ is atomic,
\item $\mathop{\textrm{atoms}}(\|_Y\cals_n) = \Union_n\mathop{\textrm{atoms}}(\cals_n)$.
\end{itemize}

The \emph{total set of criteria} of an egalitarian transition structure or rewrite system is:
\begin{itemize}
\item $\mathop{\mathrm{criteria}}(\cals) = \emptyset$ \; if $\cals$ is atomic,
\item $\mathop{\mathrm{criteria}}(\|_Y\cals_n) = \widetilde{Y} \union \Union_n\mathop{\mathrm{criteria}}(\cals_n)$,
\end{itemize}
where $\widetilde{Y}=\{\{p,q\} \mid (p,q)\in Y\}$ (so that $(p,q)$ and $(q,p)$ represent the same criterion).

Two egalitarian structures or systems $\cals_1$ and $\cals_2$ are said to be \emph{equivalent} iff $\mathop{\textrm{atoms}}(\cals_1)=\mathop{\textrm{atoms}}(\cals_2)$ and $\mathop{\mathrm{criteria}}(\cals_1)=\mathop{\mathrm{criteria}}(\cals_2)$.
\end{defn}

\begin{prop}[equivalence to composition of atoms]
Every egalitarian transition structure or rewrite system is equivalent to one of the form $\|_Y\cals_n$ where each $\cals_n$ is atomic.
\end{prop}

Namely, $\cals=\|_Y\cals_n$ is equivalent to $\|_{Y'}\mathop{\textrm{atoms}}(\cals)$, where $Y'=\{(p,q) \mid \{p,q\}\in\mathop{\textrm{criteria}}(\cals)\}$.

In our previous work~\cite{MartVM20CompSpec,Mart21Thesis} we showed that equivalent systems represent the same behavior, as given by paths and satisfaction of temporal formulas. This allows us to group the atomic components in the most suitable way for a modular design. Thus, in the example in Section~\ref{sec:river}, we first composed \lstinline!RIVER || PREVIOUS! to obtain \lstinline!RIVER-W-PREV!, which was then used in the composition \lstinline!RIVER-W-PREV || AVOID1 || AVOID2!.

In short, the compound $\|_Y\cals_n$ is a set of atomic components linked by synchronization criteria. The behavior it models is that in which each component evolves according to its internal specification, with the added restriction that all synchronization criteria have to be satisfied at all times.

\begin{defn}[signature and properties of a compound]
Let $\calr_n=(S_n,{\le}_n,\Sigma_n,E_n,R_n)\in\aers$ for $n=1,\dots,N$. Let $Y$ be a set of suitable synchronization criteria. The set of properties for $\calr_n$ has already being defined as $\Sigma|_{\intt{Stage}_n}$. The set of properties for $\|_Y\calr_n$ is defined to be $\biguplus_nP_n$. Also, the signature for $\|_Y\calr_n$ is defined to be $(\Union_nS_n,\Union_n{\le}_n,\Union_n\Sigma_n)$.
\end{defn}

This definition, as was the case for Definition~\ref{def:suitable}, is not technically precise, because we require at the same time that the namespaces be disjoint and that they share the common equational theories. A precise definition would involve pushouts. We avoid it and allow the slight informality of saying that each rewrite system is its own namespace, disjoint from the rest except for those common equational theories.

\begin{defn}[semantics in the atomic case]
Given $\calr=(S,{\le},\Sigma,E,R)\in\aers$, we define $\sem(\calr)=(Q,T,{\rew},P,g_0)\in\aets$ by:
\begin{itemize}
\setlength\itemsep{2pt}
\item
   $Q = T_{\Sigma/E,\intt{State}}$ (that is, $E$-equational classes of \lstinline!State! terms);
\item
   $T = T_{\Sigma/E,\intt{Trans}}$ (that is, $E$-equational classes of \lstinline!Trans! terms);
\item
   $\rew$ is the half-rewrite relation $\rew^{\text{eg}}_{\calr}$ induced by $R$~\cite[Definition~6]{MartVM20CompSpec};
\item
   $P=\Sigma|_\intt{Stage}$;
\item
   $g_0=[\intt{init}]_{E}$ (that is, the $E$-equational class of \lstinline!init!).
\end{itemize}
\end{defn}
The half-rewrite relation $\to$ takes the system from a state to a transition, or vice versa, in contrast to the usual state-to-state rewrites. Roughly speaking, a rewrite rule $u \trans{t} u'$ produces half rewrites from instances of $u$ to instances of $t$, and from there to instances of $u'$.

\begin{defn}[semantics for the general egalitarian case]
Given $\|_Y\calr_n\in\ers$, we define its semantics componentwise:
\[\sem(\|_Y\calr_n) = \|_Y\sem(\calr_n)\in\ets.\]
\end{defn}

A path semantics for the composition of egalitarian structures is given in Section~\ref{sec:paths}.

\subsection{Plain structures and systems}
\label{sec:plain}

In addition to egalitarian structures and systems, we use standard ones which we call \emph{plain} to avoid confusion with the egalitarian ones. An important feature of plain structures and systems is that they only have states, and not (explicit) transitions, and this allows their composition to be defined as a tuple construction. We see plain structures and systems as modeling the global behavior of composed systems, while we use egalitarian structures and systems to model local and distributed systems. The correspondence between them is given by the \emph{split} operation defined later.

\begin{defn}[plain transition structure]
A \emph{plain transition structure} is a tuple $\calt=(Q,\rew,P,q_0)$, where:
\begin{itemize}
\item
   $Q$ is the set of states;
\item
   ${\rew}\subseteq Q\times Q$ is the adjacency relation;
\item
   $P$ is the set of properties, each one a total function $p$ from $Q$ to some codomain $C_p$;
\item
   $q_0\in Q$ is the initial state.
\end{itemize}
The class of all plain transition structures is denoted by \pts.
\end{defn}

\begin{defn}[plain rewrite system]
\label{def:rs}
A \emph{plain rewrite system} is a tuple $(S,{\le},\Sigma,E,R)$, where:
\begin{itemize}
\item
   $(S,{\le})$ is a poset of sorts which contains the element \lstinline!State!.
\item
   $\Sigma$ is a signature of operators which includes the constant $\intt{init}$ of sort \lstinline!State!.
\item
   $E$ is a set of equations as in Definition~\ref{def:aers}.
\item
   $R$ is a set of rules of the form $t \to t' \rif C$, where $t,t'\in T_\Sigma(X)_s$ for some $s\in S$, and $C$ (which may be absent) is as in Definition~\ref{def:aers}.
\end{itemize}
We also refer as \emph{signature} to the triple $(S,{\le},\Sigma)$. We call \emph{properties} to the elements of $\Sigma|_\intt{State}$. The class of all plain rewrite systems is denoted by \prs.
\end{defn}

\begin{defn}[composition for plain transition structures]
\label{def:sync-comp-ts}
Given plain transition structures $\calt_n=(Q_n,{\rew}_n,\allowbreak P_n,q_{n0})\in\pts$, for $n=1,\dots,N$, their \emph{synchronous composition} with respect to the synchronization criteria $Y\subseteq \Union_nP_n\times\Union_nP_n$, is denoted by $\|_Y\{\calt_n \mid n=1,\dots,N\}$, or usually just $\|_Y\calt_n$, and is defined to be $\calt=(Q,\rew,P,q_0)\in\pts$, where:
\begin{itemize}
\setlength\itemsep{2pt}
\item
   $Q = \{\langle q_1,\dots,q_N\rangle\in\prod_nQ_n \mid \text{ for each }(p,p')\in Y \text{ with }p\in P_m \text{ and } p'\in P_{m'}\text{ we}\allowbreak \text{ have } p(q_m)=p'(q_{m'})\}$;
\item
   for $\langle q_1,\dots,q_N\rangle, \langle q'_1,\dots,q'_N\rangle\in Q$, we have $\langle q_1,\dots,q_N\rangle \rew \langle q'_1,\dots,q'_N\rangle$ iff for each $n$ either $q_n\rew_n q'_n$ or $q_n=q'_n$, with at least one occurrence of the former;
\item
   $P = \Union_nP_n$ and, if $p$ is a property originally defined in the component $\calt_m$, then it is defined in $\calt$ by $p(\langle q_1,\dots,q_N\rangle)=p(q_m)$;
\item
   $q_0=\langle q_{10},\dots,q_{N0}\rangle$, assumed to be in $Q$ (that is, to satisfy the criteria in $Y$).
\end{itemize}
\end{defn}
It is an important detail that the composition of plain transition structures can be evaluated to a single, monolithic structure of the same type, while the composition of egalitarian structures is just a set of interacting but independent components.

The composition of plain rewrite systems is defined next by a tuple-like construction; in particular, rewrite rules are produced in this way. For this to work, we need the components involved to be topmost. A plain rewrite system is said to be \emph{topmost} if its rules can only be applied on whole \lstinline!State! terms, not on its subterms---see more explanations in our previous work~\cite{MartVM20CompSpec}.

\begin{defn}[composition for plain rewrite systems]
Given plain rewrite systems $\calr_n=(S_n,{\le}_n,\Sigma_n,\allowbreak E_n,M_n,R_n)\in\prs$ for $n=1,\dots,N$, all of them topmost, their \emph{synchronous composition} with respect to synchronization criteria $Y$ is denoted by $\|_Y\{\calr_n \mid n=1,\dots,N\}$, or usually just $\|_Y\calr_n$, and is defined to be a new plain rewrite system $\calr=(S,{\le},\Sigma,E,R)\in\prs$. The elements of $\calr$ are defined as the disjoint union of the respective elements of each $\calr_n$ (that is, $S=\biguplus_nS_n$, and so on), except for the following:
\begin{itemize}
\item
   There is in $S$ a new sort \lstinline!State! and a constructor $\langle\_\rangle : \intt{State}_1\times\dots\times\intt{State}_N \pmap \intt{State}$ ($\intt{State}_n$ denotes the sort \lstinline!State! from component $\calr_n$).
\item
   There is a constant \lstinline!init! of sort \lstinline!State! and an equation $\intt{init} = \langle\intt{init}_1,\dots,\intt{init}_N\rangle$ ($\intt{init}_n$ denotes the constant \lstinline!init! from component $\calr_n$).
\item
   For each $(p,p')\in Y$, suitability of $Y$ (Definition~\ref{def:suitable}) implies the existence of a common sort $s$ and a common equational theory for it. These are common and, thus, included only once in the result of the composition.
\item
   For each property $p$ defined in the component $\calr_m$, there is in $\Sigma$ a declaration of a property with the same name and in $E$ an equation $p(\langle q_1,\dots,q_N\rangle)=p(q_m)$.
\item
   We assume an equational theory of the Booleans is included, and we add the declaration of a new operator $\intt{isValidState} : \Pi_{i=1}^N\intt{State}_i\to \intt{Boolean}$, defined by this equation:
   \[\intt{isValidState}(\langle q_1,\dots,q_N\rangle) = \Land_{(p,p')\in Y} p(\langle q_1,\dots,q_N\rangle)=p'(\langle q_1,\dots,q_N\rangle).\]
\item
   The rewrite rules from the components are dropped, and the set of rules $R$ for the composition is built in the following way. For each nonempty set $M\subseteq\{1,\dots,N\}$, and for each set of rules $q_m\rew q'_m \rif C_m$, one from each $R_m$ for $m\in M$, and setting $q'_m=q_m$ for $m\not\in M$, there is the following rule in $R$:
   \begin{align*}
   \langle q_1,\dots,q_N\rangle \rew \langle q'_1,\dots,q'_N\rangle \rif \Land_{m\in M}C_m \;&\land\; \intt{isValidState}(\langle q_1,\dots,q_N\rangle) \\
   &\land\; \intt{isValidState}(\langle q'_1,\dots,q'_N\rangle).
\end{align*}

\end{itemize}

\end{defn}
With these rules, only \lstinline!State! terms for which synchronization criteria are satisfied are reachable from \lstinline!init!.

Equations from different components are mixed together, according to this definition, but there are no conflicts, because each component is its own namespace. The resulting plain rewrite system happens to be topmost as well, so it can be used as a component in turn.

\begin{defn}[semantics for plain rewrite systems]
Given $\calr=(S,{\le},\Sigma,E,R)\in\prs$, we define its semantics $\sem(\calr)=(Q,\rew,P,q_0)\in\pts$ by:
\begin{itemize}
\setlength\itemsep{2pt}
\item
   $Q = T_{\Sigma/E,\intt{State}}$;
\item
   $\rew$ is the rewrite relation $\rew_{\calr}$ induced by $\calr$;
\item
   $P=\Sigma|_\intt{State}$;
\item
   $q_0=[\intt{init}]_{E}$.
\end{itemize}
\end{defn}

Concepts of equivalence can be defined for plain transition structures and for plain rewrite systems~\cite{MartVM20CompSpec}, corresponding to the equivalence in the egalitarian setting from Definition~\ref{def:equiv-sys}, to formalize the idea that the ordering and grouping of components in a composition are immaterial. For example, $(\cals_1\|_{Y_1}\cals_2)\|_{Y_2}\cals_3$ is equivalent to $(\cals_3\|_{Y_3}\cals_1)\|_{Y_4}\cals_2$ if $Y_1\union Y_2=Y_3\union Y_4$, for $\cals_n$ either plain rewrite systems or plain transition structures. (Remember that, whenever we write such composition expressions, we are assuming the synchronization criteria to be suitable.) Although we do not repeat those definitions here, when we write expressions like $\langle q_1,\dots,q_N\rangle\in\|_Y\cals_n$ we are assuming that some ordering and grouping of the components have been arbitrarily fixed. And when we say that two systems are equal, we rather mean they are equivalent in that sense. This is the case in the following proposition.

\begin{prop}[semantics and composition commute]
For plain rewrite systems $\calr_n$, each of them topmost, and for suitable synchronization criteria $Y$, we have that $\sem(\|_Y\calr_n)=\|_Y\sem(\calr_n)$.
\end{prop}

\subsection{The split}
\label{sec:split}

Plain systems have the advantage that they are standard rewrite systems and existing theoretical and practical tools can be used on them. For that reason, it is sometimes useful to transform an egalitarian system into an equivalent plain one. This is what the operation that we call \emph{split} does. The result of the split represents in a single system the joint evolution of the three components.

\begin{defn}[the split]
\label{def:split}
Given $\calt=(Q,T,\rew,P,g_0)\in\aets$, its \emph{split} is $\splitop(\calt)=({Q\union T},\rew,P,g_0)\in\pts$. That is, stages are transformed into states.

Given $\calr=(S,{\le},\Sigma,E,R)\in\aers$, its split is $\splitop(\calr)=(S',{\le},\Sigma,\allowbreak E,R')\in\prs$, where
\begin{itemize}
\item
   $S'$ is the result of renaming in $S$ the sort \lstinline!State! to \lstinline!State'!, and \lstinline!Stage! to \lstinline!State! (with the only aim of getting the top sort still being called \lstinline!State!), and
\item
   $R'$ is the result of splitting each rule $s \trans{t} s' \rif C$ in $R$ to produce the two rules $s \rew t \rif C$ and $t \rew s' \rif C$ in $R'$.
\end{itemize}

For a nonatomic system $\|_Y\cals_n$ in \ets\ (resp., in \ers), its split is recursively defined by $\splitop(\|_Y\cals_n) = \|_Y\splitop(\cals_n)$, a system in \pts\ (resp., in \prs).
\end{defn}

The composition of plain systems can always be evaluated to a single one, so the result of a split is always a single plain transition structure or rewrite system.

\begin{prop}[semantics and split commute]
\label{prop:sem-spl-comm}
For $\calr\in\ers$ all whose atomic components are topmost, we have that $\sem(\splitop(\calr)) = \splitop(\sem(\calr))$.
\end{prop}

\begin{defn}[compatible stages]
\label{def:comp-stage}
Given $\calt=(Q,T,\rew,P,g_0)$ and $\calt'=(Q',T',\rew',P',g'_0)$, the stages $g\in Q\union T$ and $g'\in Q'\union T'$ are said to be \emph{compatible} (with respect to $Y$) iff all criteria in $Y$ are satisfied when evaluated at them, that is, $p(g)=p'(g')$ for each $(p,p')\in Y\inters(P\times P')$. More in general, given $\calt_n=(Q_n,T_n,\rew_n,P_n,g_{n0})$ for $n=1,\dots,N$, we say that the stages $\{g_n\}_n$, with $g_n\in Q_n\union T_n$, are \emph{compatible} when they are so pairwise according to the above.
\end{defn}
   
The intuitive meaning is that compatible stages can be visited simultaneously, each within its own component system. In the example of the chained buffers, Section~\ref{sec:buffers}, the states \lstinline!sending! in \lstinline!BUFFER1! and \lstinline!receiving! in \lstinline!BUFFER2! are compatible with respect to the synchronization criterion \lstinline!BUFFER1$isSending = BUFFER2$isReceiving!, because \lstinline!isSending! evaluates to true at \lstinline!sending! and \lstinline!isReceiving! evaluates also to true at \lstinline!receiving!. There is a trivial bijection between compatible stages and states in the split which justifies the view that states in $\splitop(\calt)$ represent global states for the compound $\calt$.
   
\begin{prop}[distributed and global states]
\label{prop:bij-states}
There is a bijection between the set of compatible stages in $\|_Y\calt_n$ and the set of states in $\splitop(\|_Y\calt_n)$.
\end{prop}

\section{Distributed and global paths}
\label{sec:paths}

In preparation for the definition of satisfaction in following sections, we need an operational, or step, semantics for all our transition structures. They are given by paths (for atomic and plain structures) and sets of compatible paths (for compounds). They are defined in this section.

\begin{defn}[path and maximal path]
A \emph{path} in $\calt=(Q,T,\rew,P,g_0)\in\aets$ is a finite or infinite sequence of adjacent stages $\overline{g}=g_0\rew g_1\rew\dots$ starting at the structure's initial stage. We call such a path \emph{maximal} if it is either infinite or it is finite and its final stage has no stages adjacent to it.

Similarly, a path in $\calt=(Q,\rew,P,q_0)\in\pts$ is a sequence of adjacent states $\overline{q}=q_0\rew q_1\rew\dots$.  We call such a path \emph{maximal} if it is either infinite or it is finite and its final state has no states adjacent to it.
\end{defn}

Compatibility of paths is defined by means of a relation between indices which shows a way in which all paths can be traversed together, interleaving some steps, making other simultaneous, and keeping compatibility of stages at all times. The intuitive meaning of the following definition is that, if $\langle i_1,\dots,i_N\rangle$ is in the relation $X$, then the stages $g_{1i_1},\dots,g_{Ni_N}$ are visited at the same time, each in its structure. Thus, each relation $X$ describes a possible execution of the composed system.

\begin{defn}[compatible paths]
\label{def:compatible-paths}
Let $\calt_n\in\aets$ for $n=1,\dots,N$. For each $n$, let $\overline{g_n}=g_{n0}\rew g_{n1}\rew\dots$ be a finite or infinite path in $\calt_n$. The paths $\{\overline{g_n} \mid n=1,\dots,N\}$ are said to be \emph{compatible} (with respect to a given $Y$) iff there exists a relation between indices $X\subseteq\mathbb{N}^{\{1,\dots,N\}}$ satisfying the following conditions:
\begin{enumerate}
\item
   $\langle0,\dots,0\rangle\in X$.
\item\label{item:next}
   If $\langle i_1,\dots,i_N\rangle\in X$ and $g_{ni_n}$ is not the last stage in $\overline{g_n}$ for at least one $n\in\{1,\dots,N\}$, then for exactly one nonempty $M\subseteq\{1,\dots,N\}$ we have that $\langle i'_1,\dots,i'_N\rangle\in X$, where $i'_n=i_n+1$ if $n\in M$, and $i'_n=i_n$ otherwise.
\item
   All tuples in $X$ can be obtained by means of the two previous conditions.
\item
   $\langle i_1,\dots,i_N\rangle\in X$ implies the compatibility (with respect to $Y$) of the stages $g_{ni_n}$ ($n=1,\dots,N$).
\item\label{item:entire}
   For each stage $g_{ni}$ in each path $\overline{g_n}$, the index $i$ appears as the $n$th component of some tuple in $X$.
\end{enumerate}

Further, a set of paths is said to be \emph{maximally compatible} if no path or subset of paths in it can be extended with new stages in the respective components while maintaining compatibility.
\end{defn}

The conditions, specially Condition~\ref{item:next}, make it possible to arrange all the tuples in $X$ in a linear sequence, which is shown in Proposition~\ref{prop:bij-paths} to correspond to a path in the split system. Thus, paths in the split can be seen as global paths.

Condition~\ref{item:entire} entails that the paths are all traversed together \emph{in their entirety}. This, however, does not mean each path is maximal in its component: a partial path can be a member of a compatible set, as long as $X$ shows how to traverse it to its last (though maybe not terminal) stage.

For example, consider the paths for the chained buffers from Section~\ref{sec:buffers}
\begin{itemize}
\item
   in \lstinline!BUFFER1!: \lstinline!idle! $\to$ \lstinline!receiving! $\to$ \lstinline!gotToken! $\to$ \lstinline!sending! $\to\cdots$;
\item
   in \lstinline!BUFFER2!: the same as in \lstinline!BUFFER1!;
\item
   in \lstinline!BUFFER3!: the single-stage path \lstinline!idle!.
\end{itemize}
A set $X$ showing how to traverse these three paths would include, among others, the following triples:
\begin{itemize}
\item
   $\langle 0,0,0\rangle$, representing the three paths starting at \lstinline!idle!;
\item
   $\langle 1,0,0\rangle$ and $\langle 2,0,0\rangle$, representing only the first path advancing one step and two steps;
\item
   $\langle 3,1,0\rangle$, representing the first and second paths advancing to the respective stages \lstinline!sending! and \lstinline!receiving!, which are compatible.
\end{itemize}

As a side note, compatibility of paths cannot be defined pairwise, as we did for compatibility of stages in Definition~\ref{def:comp-stage}. It need not be the case that three paths can be traversed simultaneously keeping compatibility of stages, even if any two of them can.

Consider $\splitop(\|_Y\calt_n)$. Its states are tuples of components' stages. Thus, for each atomic component $\calt_n$ of $\calt$, a projection map $\pi_n$ can be defined from the states of $\splitop(\|_Y\calt_n)$ to the stages of $\calt_n$. This projection can be extended to paths. However, our definitions allow for a component to advance while others stay in the same stage, so, in general, a pure projection would produce repeated stages (stuttering) that we want to remove.

\begin{defn}[projection]
Let $q=\langle g_1,\dots,g_N\rangle$ be a state of $\splitop(\|_Y\calt_n)$ and $\overline{q}=q_0\rew q_1\rew\dots$ be a path in $\splitop(\|_Y\calt_n)$. For each $n=1,\dots,N$:
\begin{itemize}
\item
   we define $\pi_n(q)=g_n$;
\item
   we define $\pi_n(\overline{q})$ as the result of removing stuttering (that is, simplifying consecutive repetitions) from $\pi_n(q_0) \rew \pi_n(q_1) \rew\dots$
\end{itemize}
\end{defn}

\begin{prop}[distributed and global paths]
\label{prop:bij-paths}
There is a bijection between sets of compatible paths in $\{\calt_n\}_n$ (with respect to $Y$) and paths in $\splitop(\|_Y\calt_n)$. Also, there is a bijection between sets of maximally compatible paths in $\{\calt_n\}_n$ (with respect to $Y$) and maximal paths in $\splitop(\|_Y\calt_n)$.
\end{prop}

The paths in a compatible set are not required to be maximal. Indeed, any projection of $\overline{q}$ may fail to be maximal in its component, even if $\overline{q}$ is in the split.

\begin{proof}
We prove first that, for $\overline{q}$ a path in $\splitop(\|_Y\calt_n)$, the projections $\pi_n(\overline{q})$, for $n=1,\dots,N$, are compatible paths, with the relation $X$ (required by Definition~\ref{def:compatible-paths}) being induced by $\overline{q}$ itself. Because the projections $\pi_n$ remove stuttering, we need to be careful with the resulting indices. We introduce a function $s$ which, when applied to $\pi_n(q_i)$ (the $n$th component of the $i$th state appearing in $\overline{q}$), returns the index of that stage in the path $\pi_n(\overline{q})$ (that is, after removing stuttering). Then, $X=\{\langle s(\pi_1(q_i)),\dots,s(\pi_N(q_i))\rangle \mid q_i\text{ in }\overline{q}\}$ meets the conditions in Definition~\ref{def:compatible-paths}.

Next, we prove that, given paths $\overline{g_n}$ in $\calt_n$, for $n=1,\dots,N$, which are compatible, there is a unique path $\overline{q}$ in $\splitop(\|_Y\calt_n)$ such that $\pi_n(\overline{q})=\overline{g_n}$. Let $X$ be the relation whose existence is given by compatibility of paths in Definition~\ref{def:compatible-paths}. Let the initial state of $\overline{q}$ be $q_0=\langle g_{10},\dots,g_{N0}\rangle$. Then, inductively, for each state $q_k$ already in the path, let the next state $q_{k+1}$ be the tuple whose existence is required by Condition~\ref{item:next} in Definition~\ref{def:compatible-paths}. Condition~\ref{item:entire} ensures that the projections of this $\overline{q}$ produce the complete $\overline{g_n}$'s.

The \emph{maximal} part now follows: to any hypothetical extension for a set of compatible paths would correspond an extension to the corresponding path in the split, and vice versa.
\end{proof}

We are not saying too much here: there is an almost trivial correspondence between tuples of paths and paths of tuples. But there are useful consequences. The split provides global, monolithic concepts of states and paths. The equivalence between those concepts and the distributed ones validates our definitions and allows us to work using the most suitable view in each case. Also, as discussed in Section~\ref{sec:back-standard}, it allows us to reason about models of distributed systems, or even execute them, by performing the split and using existing techniques and tools for the corresponding global, monolithic result.

\subsection{A short diversion on locality}

Even though the definition of compatibility involves all paths at once, and thus all components at once, there is room to see locality somewhat concealed in it. We have already mentioned that the contrast between local and global, or, equivalently, between a distributed view of complex systems and a monolithic one is a motivation for our work, so a short diversion is in order.

For an example, consider a system composed of a sender and a receiver, which synchronize on a Boolean property, very much in the same way as the chained buffers in Section~\ref{sec:buffers} did:
\begin{lstlisting}
sync SENDER || RECEIVER
   on SENDER$isSending = RECEIVER$isReceiving .
\end{lstlisting}
While the \lstinline!SENDER! is not ready to send, its property \lstinline!isSending! keeps being false, the same as \lstinline!RECEIVER$isReceiving!. Meanwhile, \lstinline!SENDER! can evolve in whatever way fits to its function. The system \lstinline!SENDER! may even be a composed system on its own, and then its components can interact among them as they need to, with no concern about \lstinline!RECEIVER!. Of course, the same is true of \lstinline!RECEIVER!. This is the sense in which locality is included in our definitions. This view is more difficult to appreciate when only considering global, monolithic definitions of composition. Let us be more precise.

\begin{prop}[compatibility and locality]
Suppose given the egalitarian transition structure $\calt = \|_Y\{\calt_n \mid n=1,\dots,N\}$, which we rather prefer to view grouped as
$\calt = \calt'\|_{Y_3}\calt''$ with $\calt'= \|_{Y_1}\{\calt_n \mid n=1,\dots,N'\}$ and $\calt''=\|_{Y_2}\{\calt_n \mid n=N'+1,\dots,N\}$ (therefore, $Y=Y_1\uplus Y_2\uplus Y_3$). Suppose further that the stages $\{g_n\}_n$, $n=1,\dots,N$, are compatible and that, for each $n$, there is a $g'_n$ such that either $g_n\rew_n g'_n$ or $g_n=g'_n$ (that is, either $\calt_n$ advances one step or stays where it was). We have that the stages $\{g'_n \mid n=1,\dots,N\}$ are compatible if (but not only if) the three following conditions hold:
\begin{itemize}
\item 
   the stages in the set $\{g'_n \mid n=1,\dots,N'\}$ are compatible respect to $Y_1$;
\item 
   the stages in the set $\{g'_n \mid n=N'+1,\dots,N\}$ are compatible respect to $Y_2$; and
\item 
   for each $p$ used in $Y_3$, if $p\in P_m$, we have $p(g_m)=p(g'_m)$.
\end{itemize}
\end{prop}

\begin{proof}
Let $(p,q)$ be a criterion in $Y\inters(P_i\times P_j)$, that is, property $p$ is defined in $\calt_i$ and property $q$ in $\calt_j$. We need to show that $p(g'_i)=q(g'_j)$ if the three conditions hold.

If $i,j\in\{1,\dots,N'\}$, it means that $(p,q)\in Y_1$, and then $p(g_i)=q(g_j)$ because of the first item in the proposition statement. Similarly if $i,j\in\{N'+1,\dots,N\}$. Finally, if $i\in\{1,\dots,N'\}$ and $j\in\{N'+1,\dots,N\}$, or vice versa, then $(p,q)\in Y_3$, so that, because of the third item in the statement and the compatibility of $\{g_n\}_n$, we have $p(g'_i)=p(g_i)=q(g_j)=q(g'_j)$.
\end{proof}

\section{Linear temporal logic}
\label{sec:ltl}

The temporal logic we use in this work is \ltl~\cite[Ch.~3]{ClarGP99MC} with two deviations from the standard that we discuss below. \ltl\ is appropriate for compositional verification because its formulas are implicitly universally quantified over execution paths. Thus, when the possible executions of a system are restricted by its interaction with the environment, the remaining ones still satisfy whatever \ltl\ formulas were satisfied in isolation. ACTL*~\cite[Ch.~3]{ClarGP99MC} is a superset of \ltl\ that shares this universality property, but we restrict to LTL in this paper.

The first difference between our logic, which we call $\ltlmsp$, and standard \ltl\ is that we avoid the use of the \emph{next} temporal operator, usually represented by $\tn$ (or, alternatively, $\mathbf{N}$ or $\mathbf{X}$). The reason for avoiding $\tn$ is that its reference (the next stage) is not preserved by composition, nor by refinement. If we want to be ready for them, we should treat time as if it were dense: between the present and any \emph{next} stage, a new stage may show up. Also, the semantics for the $\tn$ operator is not clear when we have to evaluate it at both states and transitions. The resulting logic is still quite common in the literature. If we are allowed to bring in some experts to support us:
\begin{quote}
[\dots] increasing the expressiveness of our temporal logic with a \emph{next} operator would destroy the entire logical foundation for its use in hierarchical methods.~\cite{Lamp83TempLogic}
\end{quote} 

\begin{quote}
This definition is appropriate for reasoning about asynchronous processes since there is no notion of \emph{next system state} in such cases.~\cite{ClarLM89CompMC}
\end{quote}

The downside of quitting the \emph{next} operator is that, well, sometimes it is useful. In particular, in our examples, we have found often the need to specify that a formula holds at each state from the current one but excluding the current one, which in LTL would be written as $\tn\tbox\varphi$. However, we have also found that the \emph{next state of interest} can often be characterized by particular changes in the values of propositions (or, rather, properties). For example, for a proposition $p$ and a temporal formula $\varphi$, the expression $p\land(p \tu (\neg p \land \varphi))$ can be interpreted as saying that a change in the value of $p$ identifies the next state of interest, at which point we require $\varphi$ to hold. 

The second difference between $\ltlmsp$ and standard \ltl\ is that, instead of atomic propositions, we use in our formulas \emph{properties} and terms involving them---that is what $\Pi$ is for. We usually denote by $\Pi$ the set of property symbols to build formulas on, and by $P$ the set of actual properties defined in a transition structure. We decided that properties are the interfaces of systems, and that they are all that is to be observed and known from the external world. It makes sense to use them in formulas. For instance, $\tdiam(p=5)$ and $(p_1+p_2<p_3) \tu (p_4=\intt{true})$ are valid temporal formulas for us, interpretable on structures in which the respective properties, $p$, $p_i$, are defined. Using properties instead of propositions does not increase the expressive power of our formulas, because any Boolean expression involving properties can be turned into an atomic proposition (see Propositions~\ref{prop:into-standard} and~\ref{prop:plain-eq-standard}), but properties fit better in our setting.

When we get to semantics below, we will need a means to evaluate expressions involving properties. For now, from a merely syntactic point of view, we need a signature on which such expressions are built. Remember from Definitions~\ref{def:aers} and~\ref{def:rs} that a signature in rewriting logic is a triple $(S,{\le},\Sigma)$. To such a signature we add $\Pi$, a set of $S$-sorted symbols to represent properties. Then, similarly to the notations $T_\Sigma(X)$ and $T_\Sigma(X)_s$ for terms with variables from $X$, we use the notations $T_\Sigma(\Pi)$ and $T_\Sigma(\Pi)_s$ for terms which can include sorted symbols from $\Pi$. Thus, viewing such symbols as new constants, $T_\Sigma(\Pi)=T_{\Sigma\union\Pi}$ and $T_\Sigma(\Pi)_s=T_{\Sigma\union\Pi,s}$.
\begin{defn}[temporal formula]
Let $\Sigma=(S,{\le},\Sigma)$ be a signature. Let $\Pi$ be a set of $S$-sorted symbols disjoint from $\Sigma$, and let $T_\Sigma(\Pi)$ and $T_\Sigma(\Pi)_s$ be as described above. A formula in $\ltlmsp$ is defined by:
\begin{itemize}
\item $t=u$ is an atomic formula for terms $t,u\in T_\Sigma(\Pi)_s$ for some sort $s\in S$;
\item if $\varphi$ and $\psi$ are formulas, then so are $\neg\varphi$, $\varphi\lor\psi$, and $\varphi\tu\psi$.
\end{itemize}
We define $\land$, $\limpl$, $\lequiv$, $\tdiam$, $\tbox$, $\tw$, and $\tr$ as the usual abbreviations.
\end{defn}
In the particular case in which $t\in T_\Sigma(\Pi)_\intt{Bool}$ (and assuming the sort \lstinline!Bool! includes the value \lstinline!true!), it is often convenient to allow the mere $t$ as a shortcut for the formula $t=\intt{true}$, so that we can write $p_1+p_2<p_3$ instead of $(p_1+p_2<p_3)=\intt{true}$.

\section{Basic satisfaction relations}
\label{sec:basic-sat}

The satisfaction relations studied in this section consider systems as closed entities, with no environment, no interaction with other systems. Sections~\ref{sec:simul} and, specially,~\ref{sec:ag} deal with open, interacting systems.

We need two elements to jointly provide a basis to evaluate the satisfaction of $\ltlmsp$ formulas. One is a $\Sigma$-algebra on which terms in $T_{\Sigma}$ are evaluated. The other element we need is a transition structure on which temporal formulas make sense; for this, we use egalitarian transition structures and plain ones. Transition structures also provide interpretations for the property symbols in $\Pi$. Thus, we are dealing with satisfaction relations of the form $\calta\models\varphi$, where $\calt$ is a transition structure (which, in our definition, includes its initial state or stage), $\cala$ is a $\Sigma$-algebra, and $\varphi$ is a temporal formula in $\ltlmsp$.

The algebra $\cala$ is a $\Sigma$-algebra in the usual sense that it is implicitly equipped with an interpretation map for all the elements in $\Sigma$. We denote the interpretations of $s\in S$ and $f\in\Sigma$ in $\cala$, respectively, as $s_{\cala}$ and $f_{\cala}$. In the same way, a transition structure $\calt$ with set of properties $P$ is a $\Pi$-transition structure, in the sense that it is implicitly equipped with an interpretation map that assigns to each element in $\Pi$ an element in $P$. We denote the interpretation of $p\in\Pi$ in $\calt$ as $p_{\calt}$. Also, this interpretation has to be \emph{sort-preserving}, that is, if $p\in\Pi$ has been given sort $s$, then the codomain of $p_{\calt}$ has to be $s_{\cala}$. Often, $\Sigma$ and $\Pi$ are clear from context, and we omit them and say just \emph{algebra} and \emph{transition structure}.

Satisfaction is formalized below but, intuitively, evaluating $\calta\models\tbox(p_1+p_2<p_3)$ for an atomic $\calt$ entails: (i) finding the properties in $\calt$ that are the interpretations of $p_1$, $p_2$, and $p_3$; (ii) finding the values of those properties at each of $\calt$'s stages; (iii) using $\cala$ to evaluate $(p_1(g)+p_2(g)<p_3(g))=\intt{true}$ for each stage $g$; and (iv) using the results from the previous step and the adjacency relation in $\calt$ to decide whether $\tbox((p_1+p_2<p_3)=\intt{true})$ holds.

The previous discussion is equally valid for the three types of transition structures: plain or egalitarian, atomic or otherwise. The definition of satisfaction of formulas is very similar in all cases, so we present the three definitions at once, in part to avoid repetitions, but also to highlight the similarities.

Remember from Definition~\ref{def:sync-comp} that the set of properties of a composed transition structure is the disjoint union of the properties of its components. So, if each $\calt_n$ is a $\Pi_n$-transition structure, then $\calt=\|_Y\calt_n$ is a $(\biguplus_n\Pi_n)$-transition structure. The interpretation of $p$ in $\calt$, $p_\calt$, is also $p_{\calt_n}$ for some $n$.

\begin{defn}[evaluation of terms]
Consider a $\Pi$-transition structure $\calt$, with set of properties $P$, implicitly equipped with a sort-preserving interpretation for properties $p\mapsto  p_\calt$.
\begin{itemize}
\item
   For $\calt\in\aets$, consider the mapping $v : p\mapsto p_{\calt}(g_0)$, that is, the evaluation of the property $p$ at $\calt$'s initial stage.
\item 
   Respectively, for $\calt\in\ets$, consider the mapping $v : p\mapsto p_{\calt}(g_{m0})$ if $p\in P_m$, that is, the evaluation of the property $p$ at the initial stage of the component it is defined on.
\item
   Respectively, for $\calt\in\pts$, consider the mapping $v : p\mapsto p_{\calt}(q_0)$, that is, the evaluation of the property $p$ at $\calt$'s initial state.
\end{itemize}
The mapping $v$ can be extended to $T_\Sigma(\Pi)$ homomorphically in the standard way: $\overline{v}(p)=v(p)$, and $\overline{v}(f(t_1,\dots,t_n))=f_{\cala}(\overline{v}(t_1),\dots,\overline{v}(t_n))$. We denote as $t_{\calta}$ the image of $t$ under $\overline{v}$, that is, the evaluation of the term $t$ in $\calt$ and $\cala$.
\end{defn}

The type of $v(p)$ is what we called $C_p$ in Definition~\ref{def:aets}, so it is dependent on $p$.

Syntactically speaking, the role of $\Pi$ in $T_\Sigma(\Pi)$ is analogous to the role of a set of variables in $T_\Sigma(X)$. In this sense, the valuation $v$ for properties is analogous to the classical valuation maps that assign to each variable in $X$ an element in the algebra.

There is a technical point regarding interpretations and the split that we need to take care of: both $\calt$ and $\splitop(\calt)$ are $\Pi$-transition structures, both with the same set of properties, say $P$, so that they are both equipped with an interpretation from $\Pi$ to $P$, respectively, $p\mapsto p_\calt\in P$ and $p\mapsto p_{\splitop(\calt)}\in P$. In principle, the interpretations need not be the same, but that is the natural and convenient way to proceed.

\begin{defn}[the split, revisited]
\label{def:split2}
Given $\calt\in\ets$, considered as a $\Pi$-transition structure and equipped with an interpretation $p\mapsto p_\calt$, we define $\splitop(\calt)\in\pts$ as in Definition~\ref{def:split} and equipped with the interpretation $p\mapsto p_{\splitop(\calt)}=p_\calt$.
\end{defn}

\begin{defn}[$\pi^i$ and $\calt(g)$]
\begin{itemize}
\item
   For a path $\pi=\overline{g}$ in $\calt\in\aets$, we denote as $\pi^i$ the result of removing from $\pi$ its first $i$ stages. Also, $\calt(g)$ is the result of replacing in $\calt$ its initial stage by $g$, that is, $\calt(g)=(Q, T, {\rew}, P, g)$.
\item
   The definition is a little more involved for \ets. Let $\pi=\{\overline{g_n}\}_n$ be a set of compatible paths, and let $X$ be the relation from Definition~\ref{def:compatible-paths} which shows how to traverse them all simultaneously. As observed there, the tuples in $X$ can be ordered linearly. Let $\langle r_1,\dots,r_N\rangle$ be the $i$th tuple in that linear sequence. We denote as $\pi^i$ the result of removing from each component path $g_n$ its first $r_n$ stages. Also, $\calt(\{g_{nk_n}\}_n)$ is the result of replacing in each $\calt_n$ its initial stage by $g_{nk_n}$.
\item 
   Finally, for a path $\pi=\overline{q}$ in $\calt\in\pts$, we denote as $\pi^i$ the result of removing from $\pi$ its first $i$ states. Also, $\calt(q)$ is the result of replacing in $\calt$ its initial state by $q$, that is, $\calt(q)=(Q, {\rew}, P, q)$.
\end{itemize}
\end{defn}

Now, we can define the satisfaction relation for each of our three classes of transition structures.

\begin{defn}[satisfaction for transition structures]
Let $\Sigma=(S,{\le},\Sigma)$ be a signature, $\cala$ be a $\Sigma$-algebra, and $\Pi$ be a set of $S$-sorted symbols, disjoint from $\Sigma$. Also,
\begin{enumerate}
\item
   let $\calt=(Q, T, {\rew}, P, g_0)\in\aets$ be an atomic $\Pi$-structure;
\item
   respectively, let $\calt=\|_Y\calt_n\in\ets$ be a nonatomic $\Pi$-structure;
\item
  respectively, let $\calt=(Q, {\rew}, P, q_0)\in\pts$ be a plain $\Pi$-structure. 
\end{enumerate}
Finally, let $t,u\in T_\Sigma(\Pi)_s$ for some $s\in S$, and let $\varphi,\psi$ be formulas in $\ltlmsp$. The satisfaction relation $\calta\models\varphi$ is defined by:
\begin{itemize}
\item
   $\calta \models t=u$ \, iff \, $t_{\calta}=u_{\calta}$;
\item
   otherwise, $\calta \models \varphi$ iff
   \begin{enumerate}
   \item
      for each maximal path $\pi$ in $\calt$,
   \item
      respectively, for each maximally compatible set of paths $\pi$ in $\calt$,
   \item
      respectively, for each maximal path $\pi$ in $\calt$, 
   \end{enumerate}
   we have $\calta,\pi\models\varphi$.
\end{itemize}
Satisfaction of a formula by a path is defined by:
\begin{itemize}
\item
   $\calta,\pi \models t=u$ \, iff \, $\calta \models t=u$;
\item
   $\calta,\pi \models \neg\varphi$ \, iff \, not $\calta,\pi \models \varphi$;
\item
   $\calta,\pi \models \varphi\lor\psi$ \, iff \, $\calta,\pi \models \varphi$ or $\calta,\pi \models \psi$;
\item
   $\calta,\pi \models \varphi\tu\psi$ \, iff \, there is some $i\ge0$ such that $\calt(\pi_i),\cala,\pi^i \models \psi$, and for all $j<i$, we have $\calt(\pi_j),\cala,\pi^j \models \varphi$.
\end{itemize}
\end{defn}

These definitions are not only formally similar, but also equivalent in a sense made precise in the two propositions that follow. This is again an instance of the equivalence of the distributed and the monolithic views achieved through the split.

\begin{prop}[split and terms]
\label{prop:terms-eq}
For any egalitarian $\Pi$-transition structure $\calt\in\ets$, atomic or otherwise, and $\Sigma$-algebra $\cala$, we have that $t_{\calta}=t_{\splitop(\calt),\cala}$ for every term $t\in T_\Sigma(\Pi)$.
\end{prop}

\begin{proof}
The role of the structure, $\calt$ or $\splitop(\calt)$, is providing values for properties. By Definition~\ref{def:split2}, the properties of $\calt$ and those of $\splitop(\calt)$ are the same, and the interpretations are also the same.
\end{proof}

\begin{prop}[split and satisfaction for transition structures]
For any egalitarian $\Pi$-transition structure $\calt\in\ets$, atomic or otherwise, and $\Sigma$-algebra $\cala$, we have that $\calta\models\varphi$ iff $\splitop(\calt),\cala\models\varphi$ for every formula $\varphi$ in $\ltlmsp$.
\end{prop}
   
\begin{proof}
We proceed by structural induction on the shape of the formula. First:
\begin{align*}
\calta\models t=u &\iff t_\calta=u_\calta\\
&\iff t_{\splitop(\calt),\cala}=u_{\splitop(\calt),\cala}\\
&\iff \splitop(\calt),\cala\models t=u.
\end{align*}
The second equivalence is because of Proposition~\ref{prop:terms-eq}; the other two are by the definition of satisfaction.

For the inductive case, satisfaction is defined in terms of paths. We need to use the bijection between compatible sets of paths in (the atomic components of) $\calt$ and paths in its split, and between their maximal versions, from Proposition~\ref{prop:bij-paths}. We did not give a name to that bijection in the proposition, but it will be useful to have one now. For consistency, we denote by $\splitop(\pi)$ the path in $\splitop(\|_Y\calt_n)$ that corresponds to the set of paths $\pi$. (There is nothing being actually split here in the literal sense of the word, so we take it just as a convenient name.)

Then, we want to prove these equivalences:
\begin{align*}
\calta\models\varphi &\iff \text{for each } \pi \text{ max. compat. set of paths in } \calt \text{, we have } \calta,\pi\models\varphi\\
&\iff \text{for each } \pi \text{ max. path in } \splitop(\calt) \text{, we have } \splitop(\calt),\cala,\pi\models\varphi\\
&\iff \splitop(\calt),\cala\models\varphi.
\end{align*}
The middle equivalence is the one that still needs a proof. More concretely, we are going to prove something a little stronger: for each $\pi$ which is a maximally compatible set of paths in (the atomic components of) $\calt$ we have $\calta,\pi\models\varphi$ iff $\splitop(\calt),\cala,\splitop(\pi)\models\varphi$. Because each path in $\splitop(\calt)$ is the split of a set of compatible paths in $\calt$, the result follows.

We proceed again by induction on the structure of the formula. The case $t=u$ is now dealt with easily, as are negation and disjunction. Thus, what remains to be proved is: for each $\pi$ which is a maximally compatible set of paths in (the atomic components of) $\calt$, we have
\[\calta,\pi\models\varphi\tu\psi \quad\text{iff}\quad \splitop(\calt),\cala,\splitop(\pi)\models\varphi\tu\psi.\]
This chain of equivalences is quite trivial except maybe for the third one:
\begin{align*}
\calta,\pi\models\varphi\tu\psi
&\iff \exists i\ge0 \text{ such that } \calt(\pi_i),\cala,\pi^i\models\psi\\
&\qquad\qquad\text{ and } \forall j<i,\; \calt(\pi_j),\cala,\pi^j\models\varphi\\
&\iff \exists i\ge0 \text{ such that } \splitop(\calt(\pi_i)),\cala,\splitop(\pi^i)\models\psi\\
&\qquad\qquad\text{ and } \forall j<i,\; \splitop(\calt(\pi_j)),\cala,\splitop(\pi^j)\models\varphi\\
&\iff \exists i\ge0 \text{ such that } \splitop(\calt)(\splitop(\pi)_i),\cala,\splitop(\pi)^i\models\psi\\
&\qquad\qquad\text{ and } \forall j<i,\; \splitop(\calt)(\splitop(\pi)_j),\cala,\splitop(\pi)^j\models\varphi\\
&\iff \splitop(\calt),\cala,\splitop(\pi)\models\varphi\tu\psi.
\end{align*}
For the third equivalence to hold, we need $\splitop(\calt(\pi_i))=\splitop(\calt)(\splitop(\pi)_i)$, and $\splitop(\pi^i)=\splitop(\pi)^i$. Both are easy to justify, and will not be proved here.
\end{proof}

\begin{defn}[satisfaction for rewrite systems]
\label{def:sat-rs}
Let $\calr$ be an egalitarian rewrite system, atomic or otherwise, or a plain one. Let $\Sigma=(S,{\le},\Sigma)$ be its signature and $P$ be its set of properties. Let $\Pi$ be a set of $S$-sorted symbols and assume there is an interpretation from $\Pi$ to $P$. Note that $\sem(\calr)$ is a $\Pi$-transition structure. Also, let $\cala(\calr)$ be the initial algebra for the equational theory in $\calr$ (defined as the union of the equational theories of the components, if $\calr$ is not atomic). Note that $\cala(\calr)$ is a $\Sigma$-algebra. Finally, let $\varphi$ be an $\ltlmsp$ formula. We define $\calr \models \varphi$ by $\sem(\calr), \cala(\calr) \models \varphi$.
\end{defn}

\begin{prop}[split and satisfaction for rewrite systems]
\label{prop:split-sat}
In the conditions of the previous definition, let $\calr$ be an egalitarian rewrite system, and let $\varphi$ be a formula in $\ltlmsp$. We have $\calr\models\varphi$ iff $\splitop(\calr)\models\varphi$.
\end{prop}

\begin{proof}
\begin{align*}
\calr\models\varphi
&\iff \sem(\calr),\cala(\calr)\models\varphi\\
&\iff \splitop(\sem(\calr)),\cala(\calr)\models\varphi\\
&\iff \sem(\splitop(\calr)),\cala(\calr)\models\varphi\\
&\iff \sem(\splitop(\calr)),\cala(\splitop(\calr))\models\varphi\\
&\iff \splitop(\calr)\models\varphi.
\end{align*}
The third equivalence is because of Proposition~\ref{prop:sem-spl-comm}. The fourth is because $\cala(\calr)=\cala(\splitop(\calr))$, given that the equational theories from $\calr$ are copied as such into $\splitop(\calr)$.
\end{proof}

Sometimes, we say that a $\ltlmsp$ formula is a formula \emph{in the language of $\calt$} or \emph{in the language of $\calr$}, meaning that $\Sigma$ and $\Pi$ are the signature and property symbols associated with the transition structure $\calt$ or the rewrite system $\calr$, but we do not care to make $\Sigma$ and $\Pi$ explicit.

\subsection{Back to the standards}
\label{sec:back-standard}

Plain transition structures are very much like standard Kripke structures; also Boolean properties and atomic propositions are equivalent. So, in the particular case in which all properties in a plain transition structure are Boolean and all atomic formulas have the shape $p=\intt{true}$, our definitions agree with the standard ones for Kripke structures and LTL. Even out of this particular case, everything expressible in $\ltlmsp$ using properties is also expressible in LTL with Boolean propositions. And it may be worth doing so, because it would allow the use of existing tools on our nonstandard specifications. We make it formal in this section.

\begin{defn}[translation into LTL]
\label{def:form-to-prop}
We define $[\varphi]$ for any formula $\varphi\in\ltlmsp$ inductively on the structure of $\varphi$.
\begin{itemize}
\item
   From each atomic formula $t=u$ in $\ltlmsp$, we create an atomic proposition, which we denote as $[t=u]$.
\item
   For each non-atomic $\ltlmsp$ formula $\varphi$, the $\ltl$ formula $[\varphi]$ is the result of replacing each atomic subformula of $\varphi$ by its corresponding atomic proposition. That is:
   \begin{itemize}
   \item
      $[\neg\xi] = \neg[\xi]$;
   \item
      $[\xi\lor\xi'] = [\xi]\lor[\xi']$;
   \item
      $[\xi\tu\xi'] = [\xi]\tu[\xi']$.
   \end{itemize}
\end{itemize}
\end{defn}

\begin{defn}[standardization of structures]
\label{def:eg-to-kripke}
Consider given $\Sigma$, $\Pi$, and $\cala$ as usual, and an $\ltlmsp$ formula $\varphi$. Let $\calt=(Q,{\rew},P,q_0)\in\pts$. We generate a Kripke structure $\calk=\calk(\calt,\cala,\varphi)$ as $\calk=(Q,{\rew},\mathrm{AP},\call,q_0)$, where:
\begin{itemize}
\item
   $Q$, ${\rew}$, and $q_0$ are in $\calk$ the same as in $\calt$;
\item
   $\mathrm{AP} = \{[\xi]\mid\xi\textrm{ is an atomic subformula of }\varphi\}$;
\item
   $\call(q) = \{[\xi]\in\mathrm{AP} \mid \calt(q),\cala\models\xi\}$, for $q\in Q$ and $\calt(q)$ being the transition structure that results by replacing $\calt$'s initial stage by $q$.
\end{itemize}
\end{defn}

\begin{prop}[standardization of satisfaction]
\label{prop:into-standard}
Let $\Sigma$, $\Pi$, $\cala$, $\varphi$, $\calt$, $[\varphi]$, and $\calk$ as in the previous paragraphs. We have $\calta\models\varphi$ iff $\calk\models[\varphi]$.
\end{prop}

The satisfaction relation for Kripke structures and $\ltl$ formulas is the standard one~\cite{ClarGP99MC}.

\begin{proof}
The proof is an easy induction on the structure of formulas. We also need to prove the equivalence for paths: $\calta,\overline{q}\models\varphi$ iff $\calk,\overline{q}\models[\varphi]$. We illustrate it with just two cases:
\begin{align*}
\calta\models t=u
   & \iff [t=u]\in\call(q^0)\\
   & \iff \calk\models[t=u].
\end{align*}

\begin{align*}
\calta\models\varphi\tu\psi
   & \iff \textrm{for each maximal path } \overline{q} \textrm{ in } \calt \textrm{, we have } \calta,\overline{q}\models\varphi\tu\psi\\
   & \iff \textrm{for each maximal path } \overline{q} \textrm{ in } \calt, \exists i\ge0 \textrm{ such that }\\
   & \qquad\qquad \calt(q_i),\cala,\overline{q}^i\models\psi \textrm{ and } \forall j<i, \calt(q_j),\cala,\overline{q}^j\models\varphi\\
   & \iff \textrm{for each maximal path } \overline{q} \textrm{ in } \calk, \exists i\ge0 \textrm{ such that }\\
   & \qquad\qquad \calk(q_i),\overline{q}^i\models[\psi] \textrm{ and } \forall j<i, \calk(q_j),\overline{q}^j\models[\varphi]\\
   & \iff \textrm{for each maximal path } \overline{q} \textrm{ in } \calk \textrm{, we have } \calk,\overline{q}\models[\varphi]\tu[\psi]\\
   & \iff \calk\models[\varphi]\tu[\psi]\\
   & \iff \calk\models[\varphi\tu\psi].
\end{align*}
We are using here the fact that, for $\calk=\calk(\calt,\cala,\varphi)$, the Kripke structure $\calk(\pi_i)$, that is, $\calk$ with its initial state replaced by $\pi_i$, can be obtained as $\calk(\calt(\pi_i),\cala,\varphi)$, which is easy to prove.
\end{proof}

An immediate consequence is that, for $\calt\in\ets$, we have $\calta\models\varphi$ iff $\splitop(\calt),\cala\models\varphi$ iff $\calk\models[\varphi]$. This allows verifying the satisfaction of formulas in egalitarian structures by using standard tools.

It is worth noting that this procedure does not work componentwise. Suppose, for an example, that for the structure $\calt_1$ we are interested in the formula $\tbox(p_1\le5)$ for some numerical property $p_1$. In the same way, in the structure $\calt_2$ we have the formula $\tbox(p_2\ge5)$ for a numerical property $p_2$, which is to be synchronized with $p_1$. After performing the standardization procedure above on both structures, we get two Boolean propositions $[p_1\le5]$ and $[p_2\ge5]$ which are unrelated and the relation between $p_1$ and $p_2$ cannot be preserved. The implicit message is that using properties instead of Boolean propositions does not increase the expressive power of our formulas, but does increase the possibilities for synchronization.

Plain rewrite systems are close relatives of standard ones. As we have just done for transition structures, we take the final step into the standard setting.

\begin{defn}[standardization of rewrite systems]
\label{def:rplus}
From a plain rewrite system $\calr=(S,{\le},\Sigma,E,R)\in\prs$ and an $\ltlmsp$ formula $\varphi$ (for some set of property symbols $\Pi$), we define $\mathrm{AP}$ and $[\varphi]$ as in Definitions~\ref{def:form-to-prop} and~\ref{def:eg-to-kripke}. We generate a standard rewrite system $\calr(\varphi)=(S^+,{\le},\Sigma^+,E^+,R)$ in the following way. The new set of sorts $S^+$ is defined to be $S$ plus new sorts \intt{Bool} and \intt{Prop} (for atomic propositions). To obtain $\Sigma^+$ we add to $\Sigma$, for each $[\xi]\in\mathrm{AP}$, the declaration of a Boolean proposition, which we also denote by $[\xi]$, and also an operator ${\models} : \intt{State} \times \intt{Prop} \to \intt{Bool}$. Finally, we obtain $E^+$ by adding to $E$ equations to define $\models$ for each new proposition in $\mathrm{AP}$:
\begin{itemize}
\item
   $\big(g\models[t(\overline{p})=u(\overline{p})]\big) = \big(t[\overline{p}(g)/\overline{p}]=u[\overline{p}(g)/\overline{p}]\big)$, where $\overline{p}$ is the sequence of properties in $t$ (and in $u$) and $t[\overline{p}(g)/\overline{p}]$ is the result of replacing in $t$ each $p$ by its evaluation at $g$, that is, by $p(g)$.
\end{itemize}
\end{defn}
Thus, for example, $\big(g\models[p=3]\big) = \big(p(g) = 3)\big)$, assuming $p=3$ is an atomic formula in $\varphi$ for a numeric property $p$.

\begin{prop}[standardization of satisfaction]
\label{prop:plain-eq-standard}
Let $\Sigma$, $\Pi$, $\cala$, $\varphi$, $\calr$, $[\varphi]$, and $\calr(\varphi)$ as in Definition~\ref{def:rplus}. We have $\calr\models\varphi$ iff $\calr(\varphi)\models[\varphi]$.
\end{prop}

\begin{proof}
We have that $\calr\models\varphi$ \;iff\; $\sem(\calr),\cala(\calr)\models\varphi$ \;iff\; $\calk(\sem(\calr),\cala(\calr),\varphi)\models[\varphi]$ (the first equivalence by Definition~\ref{def:sat-rs}; the second by Proposition~\ref{prop:into-standard}). We assert that the Kripke structure $\calk(\sem(\calr),\cala(\calr),\varphi)$ is isomorphic to the standard semantics for rewrite systems associated to $\calr(\varphi)$. From that, the proposition follows. The assertion is not difficult to check. For example, the terms of sort \lstinline!State! are the same in $\calr$ and in $\calr(\varphi)$, and they produce state nodes in the transition structure $\sem(\calr)$, which correspond to state nodes in the Kripke structure $\calk(\sem(\calr),\cala(\calr),\varphi)$. Similarly, the adjacency relation and the values of propositions can be checked to correspond.
\end{proof}

\section{On fairness and deadlocks}
\label{sec:fair+deadlock}

When a system $\cals$ interacts with an environment $\cale$, its repertoire of execution paths is restricted to a subset of the ones that are possible when $\cals$ is run in isolation. For an LTL formula $\varphi$ in the language of $\cals$, the statement $\cals\models\varphi$ means that all maximal paths in $\cals$ satisfy $\varphi$. In particular, all maximal paths in $\cals$ that remain after the environment restriction still satisfy $\varphi$. And because $\varphi$ only speaks about $\cals$ (and not about $\cale$), we may be willing to assert that $\cals\models\varphi$ implies $\cals\|_Y\cale\models\varphi$ for any environment $\cale$ and criteria $Y$. Except this does not hold when the interaction with the environment prevents $\cals$ from executing long enough to satisfy $\varphi$. This may be the case when paths in $\cals$ that are not maximal become maximal in $\cals\|_Y\cale$, which can happen because of lack of fairness between components or because of emerging deadlocks.

Some models of interaction avoid these issues by establishing that only fair executions are part of the semantics~\cite{Pnue85TransGlobMod,GrumL94ModVerif}, or that all interactions consist of message passing and the receiver is at all times ready to receive~\cite{LyncT89IOAutom}, thus preventing emerging deadlocks. We are taking a more permissive approach, which both requires and allows a discussion of the details. This section introduces such a discussion. We consider only transition structures, but the results are equally valid for rewrite systems by means of their semantics.

First, consider deadlocks. More precisely, emerging deadlocks, that is, the ones which result from the failure of the component systems to agree on a next action to perform. 

\begin{defn}[deadlock]
\label{def:deadlock}
Let $\calt_n\in\aets$ for $n=1,\dots,N$. A set of maximally compatible paths $\{\overline{g_n}\}_n$, each $\overline{g_i}$ being a finite or infinite path in $\calt_i$, is said to be \emph{deadlocked} iff no component path $\overline{g_i}$ is maximal in its component $\calt_i$.
\end{defn}

According to this definition, there is no deadlock as long as some component system keeps running, even if only one does. Moreover, if one path in the set is finite in its system and reaches its final state in the compatible set, there is no deadlock, even though the composed system may come to a halt. This is our working definition, certainly not the only possible one. This capability to accommodate different definitions within the same framework is made possible by a permissive concept of composition like ours.

Deadlock can be prevented with some extra work on the part of the specifier. Being aware that the system is meant to work inside a largely unknown environment, the specifier should be able to anticipate unfriendly behaviors and be ready to deal with them. That is, the specification should include reactions to wrong environment behaviors, even if only with the aim of raising exceptions or performing error recovery. The following proposition shows a particular case in which this is achieved.

\begin{prop}[a case of deadlock freeness]
\label{prop:deadlock}
Let $\calt_1,\calt_2\in\aets$ and let $Y=\{(p,q)\}$ be the singleton set of synchronization criteria to compose $\calt_1$ and $\calt_2$. Suppose that in $\calt_1$ we have that, for each pair of stages $g,g'$ with $g\rew_1 g'$, and each possible value $v$ in the range of $p$, there exists a stage $g_v$, still in $\calt_1$, such that $p(g_v)=v$ and $g\rew_1 g_v$. Then no set of compatible paths in $\{\calt_1,\calt_2\}$ can be deadlocked.
\end{prop}

\begin{proof}
According to Definition~\ref{def:deadlock}, we need to prove that if $\overline{g_1}$ and $\overline{g_2}$ are compatible paths, then at least one of them is maximal in its component, $\calt_1$ or $\calt_2$. Let us consider a particular stage $g_1^1$ in $\overline{g_1}$, which is not a final stage in $\calt_1$ (if taken as an isolated system), that is, there exist $g_1^2$ with $g_1^1\rew_1 g_1^2$ in $\calt_1$. We show that $g_1^1$ cannot be the last stage in $\overline{g_1}$.

Because of the statement of the proposition, each possible value $v$ of $p$ is realized in a $g_v$ in $\calt_1$ such that $g_1^1\rew_1 g_v$. Now, suppose the relation $X$ from Definition~\ref{def:compatible-paths} (the one which shows how the two paths can be traversed) pairs $g_1^1$ with $g_2^1$, and then consider the particular value $v=q(g_2^1)$. Then, the path in $\calt_1$ that has taken us to $g_1^1$ can be extended to $g_v$ while $\calt_2$ stays at $g_2^1$, so that $\overline{g_1}$ can run indefinitely, or as long as $\calt_1$ allows it to.
\end{proof}

The conditions in Proposition~\ref{prop:deadlock} can be paraphrased as one component acting as a receiver which is ready to receive any value at any time. Less demanding conditions would be enough to guarantee absence of deadlocks.

This technique may seem too convoluted, but something similar is implicitly used in some models of composition. Typically, those models divide the possible interactions of a component with its environment into inputs and outputs. Inputs represent the reception of a value from the environment. The input value is controlled only by the environment, and the component is assumed to be ready to receive it, at any time, whatever it is. In the same way, the environment is assumed to be ready to receive any value the component outputs. For example, input/output automata~\cite{LyncT89IOAutom}, explicitly state that input events are not in control of the automata that receives them. The technique proposed in the previous paragraph is no more than an explicit implementation of this.

Consider now fairness between components. Fairness is difficult to characterize in the presence of deadlocks, so we only define it for non-deadlocked sets of paths.

\begin{defn}[fairness]
\label{def:fairness}
Let $\calt_n\in\aets$ for $n=1,\dots,N$. Consider the set of maximally compatible paths $\{\overline{g_n}\}_n$, each $\overline{g_i}$ being a finite or infinite path in $\calt_i$. Assuming there is no deadlock in $\{\overline{g_n}\}_n$, the set of compatible paths $\{\overline{g_n}\}_n$ is said to be \emph{fair} iff each component path $\overline{g_i}$ is maximal in its transition structure $\calt_i$.
\end{defn}

Thus, fairness entails that, if a partial path can be extended in its component alone, then it gets eventually extended in the composition as well. This is different from intra-component fairness: for our purposes here, we do not care about fairness inside each individual component, but only in their interactions.

Our definition of synchronous composition does not require fairness, so it is possible that a component starves. An extreme case is that in which no synchronization criteria are specified, so that the different systems are just put together, but allowed to execute independently. In this case, $\cals\models\varphi$ does not entail $\cals\|_\emptyset\cale\models\varphi$, because a possible evolution of the composed system is that $\cale$ executes but $\cals$ does not perform a single step.

A way in which fairness is ensured is by requiring synchronization infinitely often. For example, as in the following proposition.

\begin{prop}[a case of fairness]
\label{prop:sync-fair}
Let $\calt_n\in\aets$ for $n=1,\dots,N$. Suppose that for each $i,j\in\{1,\dots,N\}$, $i\ne j$, there is a pair of Boolean properties $(p,q)\in Y\inters(\calt_i\times\calt_j)$, such that \[\calt_i\models\tbox\tdiam(p=\intt{true}) \land\tbox\tdiam(p=\intt{false})\]
and
\[\calt_j\models\tbox\tdiam(q=\intt{true}) \land \tbox\tdiam(q=\intt{false}).\]
Then any set of compatible (and not deadlocked) paths in $\|_Y\calt_n$ is fair.
\end{prop}

\begin{proof}
First, we note that, for any $i=1,\dots,N$, the fact that $\calt_i \models \tbox\tdiam(p=\intt{true}) \land \tbox\tdiam(p=\intt{false})$ implies that any path in $\calt_i$ is infinite; thus, no finite maximal paths exist. According to Definition~\ref{def:fairness}, we need to prove that in any set of compatible non-deadlocked paths $\{\overline{g_n}\}_n$, each component $\overline{g_i}$ is infinite. As we are supposing they are not deadlocked, we know that at least one path is infinite. Without loss of generality, suppose $\overline{g_1}$ is infinite, let $\overline{g_i}$ be any other path, and let $q$ be the property in $\calt_i$ which synchronizes with $p$, according to the proposition's statement.

Let $X$ be the relation from Definition~\ref{def:compatible-paths} which shows how to traverse the compatible paths. Let us say $g_1^1$ and $g_i^1$ are compatible stages in $\calt_1$ and $\calt_i$, resp., appearing in $\overline{g_1}$ and $\overline{g_i}$, resp., which are paired by $X$. Further, suppose, again without loss of generality, that $p(g_1^1)=q(g_i^1)=\intt{true}$. Because $\overline{g_1}\models\tbox\tdiam(p=\intt{false})$, there is a stage $g_1^2$ in $\overline{g_1}$ which does not satisfy $p$. Therefore, so that the criterion $(p,q)$ is kept, $\overline{g_i}$ must contain a stage $g_i^2$ which does not satisfy $q$ and that is accessible from $g_i^1$, that is, $g_i^1\rew\cdots\rew g_i^2$. Thus, $\overline{g_i}$ is infinite.
\end{proof}

The condition in Proposition~\ref{prop:sync-fair} is an example. It is nice in that it can be expressed as a temporal-logic formula and, thus, checked by the usual means. More general and easy to meet conditions may be found which are sufficient to ensure fairness. As mentioned above, in some models of computation and composition, fairness is included from the start, that is, the path semantics of a specification includes, by definition, only fair executions, even though the specification, textually taken, would allow unfair ones. This is different from our view, in which we require the specification to be fair as given. The two views, however, are not completely disjoint. In Section~\ref{sec:ag}, we consider the assume/guarantee technique and mention that temporal formulas expressing fairness requirements can be added to the \emph{assume} part of a specification. This, in a sense, makes fair semantics a particular case of our model. Also, sometimes a system can be externally controlled to allow only fair executions in it. Maybe, even, such a control can be exerted via a synchronous composition with a suitable system. But many different concepts of fairness are possible, and it is not to be expected that all of them can be dealt with in this way.

\begin{prop}[deadlock freeness and fairness are enough]
\label{prop:long-enough}
Given $\|_Y\calt_n$, with $\calt_n\in\aets$ for $n=1,\dots,N$, if all sets of maximally compatible paths are non-deadlocked and fair, then $\calt_i\models\varphi$ implies $\|_Y\calt_n\models\varphi$ for each $i\in\{1,\dots,N\}$ and each formula $\varphi$ in the language of $\calt_i$.
\end{prop}

\begin{proof}
The assertion $\calt_i\models\varphi$ means that all paths in $\calt_i$ satisfy $\varphi$. Each compatible set of paths in $\|_Y\calt_n$ contains as $i$th component a path in $\calt_i$ which, because of fairness and absence of deadlock, is guaranteed to be maximal in its component system $\calt_i$. And because $\varphi$ is expressed in the language of $\calt_i$, its satisfaction does not depend on other component paths. Therefore, $\|_Y\calt_n\models\varphi$.
\end{proof}

Besides, deadlocks and fairness become unimportant when $\varphi$ is a safety formula: by definition, a safety formula is satisfied by a path iff it is satisfied by every initial segment of that path, even the empty one. Thus, the proof of the following proposition is immediate.

\begin{prop}[safety formulas are enough]
\label{prop:safety}
Given $\|_Y\calt_n$, with $\calt_n\in\aets$ for $n=1,\dots,N$, if $\varphi$ is a safety formula in the language of $\calt_i$ for some $i\in\{1,\dots,N\}$, then $\calt_i\models\varphi$ implies $\|_Y\calt_n\models\varphi$.
\end{prop}

A component from which we only require to satisfy safety formulas can be seen as imposing its behavior on the compound. It acts as a controller or a strategy. This is the case for the mutual exclusion controller example from Section~\ref{sec:mutex} (revisited in Sections~\ref{sec:mutex-cont} and~\ref{sec:mutex-cont2}). 

The next proposition is a simple remark that a kind of converse implication always holds.
\begin{prop}[satisfaction in any environment]
With the usual notation, we have that if $(\calt\|_Y\cale),\cala\models\varphi$ for every environment $\cale$ and every suitable $Y$, then $\calta\models\varphi$. Likewise, if $\calr\|_Y\cale\models\varphi$ for every environment $\cale$ and every suitable $Y$, then $\calr\models\varphi$.
\end{prop}

\begin{proof}
We can define an environment $\cale_0$ that preserves all the behaviors of $\calt$ in the following way: let $\cale_0$ consist of a single system with a unique state, and let $Y=\emptyset$, that is, no requirements for synchronization. The behaviors of $\calt$ in this environment are the same as the ones of $\calt$ alone.

We are assuming $(\calt\|_Y\cale),\cala\models\varphi$ for all $\cale$ and $Y$, so, in particular, $(\calt\|_\emptyset\cale_0),\cala\models\varphi$ and, because of the previous paragraph, also $\calta\models\varphi$. The proof for rewrite systems follows easily.
\end{proof}

Concerning deadlocks and fairness, our framework sets the responsibility in the hands of the user. This is also the case in \cite[Chapter~16]{GlabVH2019Fair}, where some actions may need to be explicitly declared as \emph{non-blocking}. And in~\cite{KlaiHI2005ModVerifPN}, where an algorithm is provided to check that an interaction between Petri nets is \emph{non-constraining}, which is a similar concept. This is another point where an actual implementation may include tools to help. We do not go deeper into this issue here.

\section{Componentwise simulation and abstraction}
\label{sec:simul}

A way to analyze a system is to find another one that in some sense behaves the same and is simpler. This is formalized with the concept of \emph{simulation}~\cite{ClarGP99MC}. A particular kind of simulation is \emph{abstraction}, in which the simpler system is obtained by forgetting some features from the original. In rewriting logic, a well studied kind of abstraction is \emph{equational abstraction}~\cite{MesePM08EqAbstr}. In this section, we show that componentwise simulation and equational abstraction translate into global ones. That is, roughly speaking, if there is a simulation (resp., equational abstraction) between systems $\cals_1$ and $\cals'_1$, then there is also a simulation (resp. equational abstraction) between $\cals_1\|_Y\cals_2$ and $\cals'_1\|_Y\cals_2$. This is sometimes phrased as simulation and equational abstraction being congruences.

Another kind of abstraction that has been studied in relation to rewriting logic is \emph{predicate abstraction}~\cite{BaeM14PredAbstr}. According to it, states of the original system which coincide in the values assigned to all atomic propositions are identified in the abstract system. Predicate abstraction in one component does not need to map to predicate abstraction in the composition. However, predicate abstraction induces a simulation in the abstract component, which does map to a simulation at the global level. We have not much else to say about predicate abstraction in this paper, though we use it in the example in Section~\ref{sec:mutex-cont}.

There are several ways in which abstraction can be useful for compositional verification. First, instead of verifying $\cals$ in the environment $\cale$ (that is, $\cals\|_Y\cale$), we can verify an abstraction of $\cals$ in the same environment. Second, if we verify $\cals$ in $\cale$, the result will also hold for any environment of which $\cale$ is an abstraction. Often, we model intuitively our systems from scratch as abstractions. This is certainly the case for the example on chained buffers in Section~\ref{sec:buffers}. The results which follow in this section show that, if we later need to refine our initial specification, verification may not need to be redone.

\subsection{Simulation}

Up to now, we have been taking care of defining each concept in both the distributed and the monolithic view. For example, we defined a compatible set of paths and showed it equivalent to a path in the split; and we then defined satisfaction of formulas based on both and, again, showed equivalence. Definition~\ref{def:simul} just below, however, defines simulation for composed egalitarian structures as simulations for their splits. We proceed in this way from now on, because it makes definitions and results easier. Still, when we want to enforce one or the other view, distributed or monolithic, we use one or the other of the two equivalent notations, like, for example, either $\|_Y\cals_n\models\varphi$ or $\splitop(\|_Y\cals_n)\models\varphi$.

\begin{defn}[simulation]
\label{def:simul}
Given a set $\Pi$ of property symbols and two atomic egalitarian $\Pi$-transition structures $\calt=(Q,T,{\rew},P,g_0)$ and $\calt'=(Q',T',{\rew}',P',g'_0)$, a \emph{simulation} ${\simul}:\calt\to\calt'$ is a relation ${\simul}\subseteq(Q\union T)\times(Q'\union T')$ such that:
\begin{itemize}
\item
   $g_0\simul g'_0$;
\item
   if $g\simul g'$ then $p_{\calt}(g)=p_{\calt'}(g')$ for each $p\in\Pi$;
\item
   if $g_1\simul g'_1$ and $g_1\rew g_2$ in $\calt$, then there exists a finite path in $\calt'$, $g'_1\rew'\dots\rew'g'_k$, with $k\ge1$, such that $g_1\simul g'_i$ for $i=1,\dots,k-1$ and $g_2\simul g'_k$.
\end{itemize}
If both $\simul$ and $\simul^{-1}$ are simulations, we say that $\simul$ is a bisimulation.

The definition for plain transition structures is a straightforward adaptation of the above. Finally, a simulation between nonatomic egalitarian transition structures $\|_Y\calt_n$ and $\|_{Y'}\calt'_n$ is, by definition, a simulation between their splits: ${\simul} : \splitop(\|_Y\calt_n) \rightarrow \splitop(\|_{Y'}\calt'_n)$.

A (bi)simulation is with respect to the symbols in $\Pi$. When we need to make this explicit, we say it is a $\Pi$-(bi)simulation.
\end{defn}

The third item in the definition allows, in particular, $k=1$, so that the requirement becomes $g_1\simul g'_1$ and $g_2\simul g'_1$---so to speak, $\calt$ advances while $\calt'$ waits.

The concept defined above is analogous to the ones called \emph{stuttering (bi)sim\-u\-la\-tion} and \emph{weak (bi)simulation} in the literature. However, we decided to avoid the use of the \emph{next} operator in our temporal logic, and also decided that only the values of properties are important, not paying attention to possible internal steps. Thus, we are always working in a way that pretty much corresponds to stuttering or weakness. So, we drop adjectives and call our concept just \emph{(bi)simulation}.

\begin{thm}[simulation and satisfaction]
\label{thm:simul-form}
Consider $\Sigma$, $\Pi$, and $\cala$ as usual, and $\calt,\calt'\in\ets\union\pts$. If there exists a simulation ${\simul}:\calt\to\calt'$, then for every $\ltlmsp$ formula $\varphi$ we have that $\calt',\cala\models\varphi$ implies $\calta\models\varphi$. If $\simul$ is a bisimulation, then $\calta\models\varphi$ iff $\calt',\cala\models\varphi$.
\end{thm}

\begin{proof}
It is an easy adaptation of the proof for more traditional settings~\cite{ClarGP99MC}. It proceeds by induction on the structure of $\varphi$. It relies on two lemmas that hold whenever there is a simulation ${\simul}:\calt\to\calt'$ (they are easy, and we do not prove them here): first, that $t_\calta = t_{\calt',\cala}$ for any term $t\in T_\Sigma(\Pi)$; second, that for each path in $\calt$ there is a (stuttering, weak) corresponding path in $\calt'$. Let us sketch just one base case and one inductive case:
\begin{align*}
\calt',\cala\models t=u
   &\iff t_{\calt',\cala}=u_{\calt',\cala}\\
   &\iff t_{\calta}=u_{\calta}\\
   &\iff \calta\models t=u.
\end{align*}
The second equivalence is justified by the first lemma mentioned above.

\begin{align*}
\calt',\cala\models\varphi\tu\psi
   & \iff \textrm{for each path } \overline{g}' \textrm{ in } \calt' \textrm{ we have } \calt',\cala,\overline{g}'\models\varphi\tu\psi\\
   & \iff \textrm{for each path } \overline{g}' \textrm{ in } \calt' \textrm{ there exists } i'\ge0 \textrm{ such that }\\
   & \qquad\quad \calt'(g'_{i'}),\cala,\overline{g}'^{i'}\models\psi \textrm{ and, for all } j'<i', \calt'(g'_{j'}),\cala,\overline{g}'^{j'}\models\varphi\\
   & \implies \textrm{for each path } \overline{g} \textrm{ in } \calt \textrm{ there exists } i\ge0 \textrm{ such that }\\
   & \qquad\quad \calt(g_i),\cala,\overline{g}^i\models\psi \textrm{ and, for all } j<i, \calt(g_j),\cala,\overline{g}^j\models\varphi\\
   & \iff \textrm{for each path } \overline{g} \textrm{ in } \calt \textrm{ we have } \calta,\overline{g}\models\varphi\tu\psi\\
   & \iff \calta\models\varphi\tu\psi.
\end{align*}
The ``$\implies$'' step in the middle is justified by the second lemma mentioned above.
\end{proof}

The next theorem is our main result about simulations, stating that componentwise simulations induce global ones. It can be seen as an adaptation of~\cite[Ch.~12]{ClarGP99MC}.

\begin{defn}[$\sim$ for synchronization criteria]
For $n=1,\dots,N$, let \[A_n=(Q_{A_n},T_{A_n},{\rew_{A_n}},P_{A_n},g_{A_n}^0) \quad\text{and}\quad B_n=(Q_{B_n},T_{B_n},{\rew_{B_n}},P_{B_n},g_{B_n}^0)\]
be atomic egalitarian $\Pi_n$-transition structures such that there are $\Pi_n$-simulations ${\simul}_n : A_n\to B_n$. Consider the composed systems $\|_Y A_n$ and $\|_Z B_n$. We denote by $Y\sim Z$ the fact that, for $p,q\in\Union_n\Pi_n$, we have $(p_{A_n},q_{A_m})\in Y\inters(P_{A_n}\times P_{A_m})$ iff $(p_{B_n},q_{B_m})\in Z\inters(P_{B_n}\times P_{B_m})$.
\end{defn}

\begin{thm}[simulation and composition]
\label{thm:comp-simul}
Let $A_n$ and $B_n$ be atomic egalitarian $\Pi_n$-transition structures such that there are $\Pi_n$-simulations ${\simul}_n : A_n\to B_n$ for $n=1,\dots,N$. (The identity is a bisimulation, so this includes the case that $A_n=B_n$ for some or all $n$.) Consider $A=\splitop(\|_Y A_n)$ and $B=\splitop(\|_Z B_n)$ for some $Y$ and $Z$ with $Y\sim Z$. Then, there is a simulation ${\simul} : A \to B$ (as plain transition structures). In addition, if all $\simul_n$ are bisimulations, then $\simul$ can be taken to be a bisimulation as well.
\end{thm}

\begin{proof}
The simulation $\simul$ is defined by
\[\langle g_{A_1},\dots,g_{A_N}\rangle \simul \langle g_{B_1},\dots,g_{B_N}\rangle \iff g_{A_n} \simul_n g_{B_n} \textrm{ for all }n.\]
We must show that this is indeed a simulation (if each $\simul_n$ is), that is, that the three items in Definition~\ref{def:simul} hold.

The first item in the definition, that $\langle g_{A_10},\dots,g_{A_N0}\rangle \simul \langle g_{B_10},\dots,g_{B_N0}\rangle$, follows immediately from $g_{A_n0} \simul_n g_{B_n0}$ holding for each $n$.

For the second item in Definition~\ref{def:simul}, we must prove that, for arbitrary $g_{A_n}$ and $g_{B_n}$, if $\langle g_{A_1},\dots,g_{A_N}\rangle \simul \langle g_{B_1},\dots,g_{B_N}\rangle$, we have $p_A(\langle g_{A_1},\dots,g_{A_N}\rangle)=p_B(\langle g_{B_1},\dots,g_{B_N}\rangle)$. So, take a particular $p_A\in\Union_n P_{A_n}$. Suppose $p_A=p_{A_k}\in P_{A_k}$ and, therefore, because $Y\sim Z$, $p_B=p_{B_k}\in P_{B_k}$. Then, $p_A(\langle g_{A_1},\dots,g_{A_N}\rangle) = p_{A_k}(g_{A_k}) = p_{B_k}(g_{B_k}) = p_B(\langle g_{B_1},\dots,g_{B_N}\rangle)$.

For the third item in Definition~\ref{def:simul}, we consider the simpler case with only two components, that is, $N=2$. This simplifies the presentation. The case for a general $N$ follows the same lines.

Thus, from $\langle g_{A_1},g_{A_2}\rangle \simul \langle g_{B_1},g_{B_2}\rangle$ and $\langle g_{A_1},g_{A_2}\rangle \rew_A \langle g'_{A_1},g'_{A_2}\rangle$ we must be able to produce a path in $B$ with the needed properties. From $\langle g_{A_1},g_{A_2}\rangle \simul \langle g_{B_1},g_{B_2}\rangle$ we get $g_{A_1} \simul_1  g_{B_1}$ and $g_{A_2} \simul_2  g_{B_2}$. And from $\langle g_{A_1},g_{A_2}\rangle \rew_A \langle g'_{A_1},g'_{A_2}\rangle$ we deduce both $(g_{A_1} \rew_{A_1} g'_{A_1}$ or $g_{A_1} = g'_{A_1})$ and $(g_{A_2} \rew_{A_2} g'_{A_2}$ or $g_{A_2} = g'_{A_2})$.

For $A_1$, if $g_{A_1} \rew_{A_1} g'_{A_1}$, because $\simul_1$ is a simulation, we have that there exist a finite path $g_{B_1}=g_{B_1}^1 \rew_{B_1} \dots \rew_{B_1} g_{B_1}^{i_1}$, $i_1\ge1$, such that $g_{A_1}\simul_1 g^1_{B_1}$, \dots, $g_{A_1}\simul_1 g_{B_1}^{i_1-1}$ and $g'_{A_1}\simul_1 g_{B_1}^{i_1}$. If instead $g_{A_1}=g'_{A_1}$, we choose $g_{B_1}=g'_{B_1}$, which can be seen as a path of length $1$. The same can be done for $A_2$, after which we end with a path in $B_1$ and another in $B_2$.

From these paths in $B_1$ and in $B_2$ we build now one in $B=B_1\|_ZB_2$. The idea is to interleave in whichever way the paths $g^1_{B_1}\rew_{B_1}^*g^{i_1-1}_{B_1}$ and $g^2_{B_2}\rew_{B_2}^*g^{i_2-2}_{B_2}$, and then take a last joint step $\langle g^{i_1-1}_{B_1}, g^{i_2-1}_{B_2}\rangle \rew_B \langle g^{i_1}_{B_1}, g^{i_2}_{B_2}\rangle$. For example:
$\langle g_{B_1},g_{B_2}\rangle=\langle g_{B_1}^1,g_{B_2}^1\rangle \rew_B^*
\langle g_{B_1}^{i_1-1},g_{B_2}^1\rangle
\rew_B^*
\langle g_{B_1}^{i_1-1},g_{B_2}^{i_2-1}\rangle
\rew_B
\langle g_{B_1}^{i_1},g_{B_2}^{i_2}\rangle$

Two points remain to be proved. First, that $\langle g_{A_1},g_{A_2}\rangle \simul g$ for all stages $g$ in the path, except the last one, and that $\langle g'_{A_1},g'_{A_2}\rangle \simul \langle g_{B_1}^{i_1},g_{B_2}^{i_2}\rangle$. This is immediate, because it holds componentwise. Second, that the exhibited path is indeed a path in $B$, that is, that all stages in it satisfy the synchronization criteria in $Z$. The key here is that stages related by the simulation assign equal values to corresponding properties. For example, for the final stage, we know that $g'_{A_1} \simul_i g^{i_1}_{B_1}$ and $g'_{A_2} \simul_i g^{i_2}_{B_2}$ and, therefore, for each property $p$ we have $p_{A_1}(g'_{A_1}) = p_{B_1}(g^{i_1}_{B_1})$ and $p_{A_2}(g'_{A_2}) = p_{B_2}(g^{i_2}_{B_2})$. But $\langle g'_{A_1}, g'_{A_2}\rangle$ is a stage in $A$, and, thus, satisfies all criteria in $Y$. Finally, because $Y\sim Z$, the criteria in $Z$ are satisfied by $\langle g^{i_1}_{B_1}, g^{i_2}_{B_2}\rangle$.
\end{proof}

On the other hand, similarly behaved systems can be specified from quite different components, so it is not to be expected that any (bi)simulation $\mathsf{S} : \mathrm{split}(T_1) \to \mathrm{split}(T_2)$, for $T_1$, $T_2$ egalitarian transition structures, can be factored as a set of (bi)simulations on the components.

Given the importance of deadlocks and fairness in our setting, as discussed in Section~\ref{sec:fair+deadlock}, it is necessary to explore how they relate to simulation. It is not difficult to see that a mere simulation does not even preserve maximal compatibility of paths, which is needed to make sense of the definitions. The situation is different with bisimulation.

\begin{prop}[bisimulations preserve fairness and deadlock-freeness]
Let $\calt_n,\calt'_n\in\aets$ for $n=1,\dots,N$, with each $\calt_n$ and $\calt'_n$ being a $\Pi_n$-transition structure. Let $Y$ be any suitable set of synchronization criteria. For convenience, we say that $\|_Y\calt_n$ (resp., $\|_Y\calt'_n$) is deadlock-free iff no set of maximally compatible paths in it is deadlocked (as defined in Definition~\ref{def:deadlock}). Similarly, we say that $\|_Y\calt_n$ (resp., $\|_Y\calt'_n$) is fair iff all non-deadlocked and maximally compatible sets of paths are fair (as defined in Definition~\ref{def:fairness}). Suppose there are bisimulations $\simul_n:\calt_n\to\calt'_n$ for each $n$. Then, $\|_Y\calt_n$ is deadlock-free iff $\|_Y\calt'_n$ is. Also, $\|_Y\calt_n$ is fair iff $\|_Y\calt'_n$ is.
\end{prop}

\begin{proof}
Given a path $\overline{g'_n}$ in $\calt'_n$, the bisimulation $\simul_n$ allows us to find a corresponding path in $\calt_n$ which we denote as $\simul_n^{-1}(\overline{g'_n})$. It is easily justified that $\simul_n^{-1}(\overline{g'_n})$ is maximal in $\calt_n$ iff $\overline{g'_n}$ is in $\calt'_n$. Also, the set of paths $\{\overline{g'_n}\}_n$ is (maximally) compatible iff the set of paths $\{\simul_n^{-1}(\overline{g'_n})\}_n$ is. The definitions of fairness and deadlock depend only on the concepts of maximal path and of maximally compatible set of paths, hence the proposition.
\end{proof}

\subsection{Equational abstraction}
\label{sec:eq-abstr}

A well-known way to implement simulations is by equational abstraction in a rewrite system~\cite{MesePM08EqAbstr}. In short, on an atomic egalitarian or plain rewrite system $\calr=(S,{\le},\Sigma,E,R)$, we can perform equational abstraction by adding equations $E'$ to obtain the new system $\calr'=(S,{\le},\Sigma,{E\union E'},R)$, so that states satisfying certain conditions are now equated and considered the same. The usual questions about computability apply here, that is, we must ensure that the new set of equations (oriented left to right) is ground Church-Rosser and terminating, and that the rules are still ground coherent with respect to the new set of equations. In~\cite[Section~3.5]{MartVM20CompSpec}, we justified that checking for computability can be made componentwise. Therefore, checking whether a global abstraction is executable can also be done componentwise.

\begin{prop}[equational abstraction induces bisimulation]
Let $\calr'=(S,{\le},\Sigma,{E\union E'},R)\in\aers$ (resp., $\prs$) be an equational abstraction of $\calr=(S,{\le},\Sigma,E,R)$. The relation $\{([t]_E,[t]_{E\union E'}) \mid t\in T_{\Sigma,\intt{Stage}}\}$ (resp., $t\in T_{\Sigma,\intt{State}}$) is a bisimulation.
\end{prop}

\begin{proof}
We have to check that the three conditions in Definition~\ref{def:simul} hold in both directions.
\begin{itemize}
\item
   Each stage (resp., state) is trivially related to its abstraction. In particular, initial ones are, which ensures the first condition is met.
\item
   The set $E$ of equations includes the ones that define the values of properties. Thus, if the extended set of equations $E\union E'$ is Church-Rosser, we infer that properties are \emph{preserved}, that is, $t\equiv_{E\union E'}u \implies p(t)=p(u)$, or, in words, that all stages (resp., states) that have been fused into the same abstract stage (resp., state) assign the same values to properties. This ensures the second condition is met.
\item
   All transitions are kept through equational abstraction. Even if two stages (resp., states) $t$ and $u$ for which $[t]_E\rew[u]_E$ get abstracted into the same, that is, $[t]_{E\union E'}=[u]_{E\union E'}$, we will still have $[t]_{E\union E'}\rew[u]_{E\union E'}$. And every transition in the abstracted system derives from one in the original one. This ensures the third condition is met.
\end{itemize}
\end{proof}

\begin{thm}[equational abstraction and composition]
\label{thm:comp-eq-abstr}
Let $\calr_n,\calr'_n\in\aers$ be such that each $\calr'_n$ is an equational abstraction of the corresponding $\calr_n$ for $n=1,\dots,N$. Consider $\|_Y\calr_n$ and $\|_Y\calr'_n$ for some set of synchronization criteria $Y$. Then, $\splitop(\|_Y\calr'_n)$ can be obtained as an equational abstraction of $\splitop(\|_Y\calr_n)$.
\end{thm}

\begin{proof}
The difference between the contribution of each $\calr_n$ to $\splitop(\|_Y\calr_n)$ and the contribution of $\calr'_n$ to $\splitop(\|_Y\calr'_n)$ are some equations. So, $\splitop(\|_Y\calr'_n)$ is $\splitop(\|_Y\calr_n)$ plus some equations, that is, an equational abstraction.
\end{proof}

\subsection{Example: mutual exclusion (continued)}
\label{sec:mutex-cont}

We apply now simulation to our mutual exclusion example from Section~\ref{sec:mutex}. For a reminder, these were the instructions we used in the specification of each of the trains:
\begin{lstlisting}
crl atStation N =[ comingFrom N ]=> atStation (N + 1) if N < 2 .
rl atStation 2 =[ crossing ]=> atStation 0 .
eq isCrossing @ crossing = true .
eq isCrossing @ G = false [owise] .
\end{lstlisting}
The property \lstinline!isCrossing! embodies all our model cares about in each train system, and it makes sense to perform abstraction based on it, so that all stages with the same value for that property get equated. In this case, equational abstraction would result in all stages except \lstinline!crossing! being equationally reduced to one of them. Equivalently, we can perform predicate abstraction to get one state for the truth of \lstinline!isCrossing! and another for its falsehood, producing the following:
\begin{lstlisting}
rl false =[ true ]=> false .
eq isCrossing @ B = B .
\end{lstlisting}
The state \lstinline!true! represents the former \lstinline!crossing!, while \lstinline!false! is the abstraction for all the other states. We call the two systems with this specification \lstinline!S-TRAIN1! and \lstinline!S-TRAIN2!.

It is quite straightforward to see that the conditions in Definition~\ref{def:simul} are met and these are indeed simulations. Because of Theorem~\ref{thm:comp-eq-abstr}, we can use this specification instead of the original one in composed systems and draw conclusions based on it.

Now we perform a three-way synchronous composition to build a new system that we call \lstinline!SAFE-TRAINS!:
\begin{lstlisting}
sync S-TRAIN1 || S-TRAIN2 || MUTEX
   on S-TRAIN1$isCrossing = MUTEX$isGranting(1)
   /\ S-TRAIN2$isCrossing = MUTEX$isGranting(2) .
\end{lstlisting}

We want to show that mutual exclusion holds for the crossings, that is:
\begin{equation}
\label{eq:mutex} \intt{SAFE-TRAINS}\models\tbox\neg(\intt{S-TRAIN1\textcolor{red!50!black}{\$}isCrossing}\land\intt{S-TRAIN2\textcolor{red!50!black}{\$}isCrossing})
\end{equation}
from which we can readily deduce the same formula holds for \lstinline!TRAIN1! and \lstinline!TRAIN2! and the same \lstinline!MUTEX!. One way to prove (\ref{eq:mutex}) is to use our prototype implementation to perform the split on \lstinline!SAFE-TRAINS! and then use Maude's model checker. A more compositional way is also possible, which is shown later, in Section~\ref{sec:mutex-cont2}.

\section{The assume/guarantee technique}
\label{sec:ag}

The classical satisfaction relation between a system $\cals$ and a temporal formula $\varphi$, which we write $\cals\models\varphi$, considers the system as if run in isolation---as a non-interacting, closed system. For open systems, techniques have been devised to verify that a component satisfies a given specification in a suitable environment. Well-known among such techniques is assume/guarantee~\cite{Pnue85TransGlobMod}, A/G from now on. This section is devoted to discussing this technique and its adaptation to our setting for verifying rewrite systems.

Satisfaction, according to the A/G technique, involves two formulas: one stating what can be assumed from the environment; the other stating what one particular component is ready to guarantee based on the assumption and on its own internal behavior. The notation we are using is $\cals\models\alpha\grt\gamma$~\cite{ElkaG+18CircAG} (or $\calta\models\alpha\grt\gamma$ for transition structures, or $\calr\models\alpha\grt\gamma$ for rewrite systems), where $\alpha$ is the assumption and $\gamma$ the guarantee. Both are $\ltlmsp$ formulas expressed in the language of $\cals$. Thus, $\alpha$ speaks about the environment by means of the properties of $\cals$, some of which are to be synchronized with the ones of the environment.

The naïve reading of $\cals\models\alpha\grt\gamma$ as ``$\cals$ guarantees the satisfaction of $\gamma$ if placed in an environment that satisfies $\alpha$'' is misleading. It is not really necessary that the environment satisfies $\alpha$---it is the interaction that matters. For example, an environment that behaves according to the CCS expression $a.P\,|\,b.Q$ does not ensure the execution of $a$ in general, but for some processes, like $a.P'\,|\,c.Q'$, it does: the environments $a.P\,|\,b.Q$ and $a.P$ are equivalent for that process, they induce the same restrictions. More in general, it is not necessary that the environment satisfies $\alpha$, but only that the interaction of the environment with the process does.

Moreover, the assumption $\alpha$ can sometimes be intuitively thought of as reflecting other convenient laws or facts, not always expected to be realized by an environment, like fairness assumptions, or the fact that time is strictly increasing in the case of timed systems~\cite{AbadL95Conjoin}. In those cases, the assumption would only involve properties not used for synchronization, so that it restricts the component and not the environment.

A definition of A/G satisfaction based on the intuitions in the previous paragraphs may consider execution paths in their fullness, that is, they may be ultimately based on assertions like ``if some full execution path satisfies $\alpha$, then it also satisfies $\gamma$''. This is unsuitable, however, because it allows that first a system fails to satisfy $\gamma$ and only later the environment fails to satisfy $\alpha$. This is probably not what we have in mind when we think about A/G. Instead, we can choose an inductive definition~\cite{MisrC81ProofNetwork,JonsY96AGinLTL}, which could be stated in this way:
\[\textrm{if }\cals\|_Y\cale\models_i\alpha\textrm{, then }\cals\models_{i+1}\gamma,\]
where $\models_i$ represents the satisfaction up to $i$ steps away from the current state.

\emph{A posteriori}, the two concepts, the full-path one and the inductive one, turn out to be equivalent, which reflects the fact that a system could only take advantage of the difference if it knew that the environment was going to fail to satisfy $\alpha$ in the future, which it cannot. Similar results are known in other settings~\cite[Theorem~5.1]{KupfV00AutModMC}~\cite[Section~5.1]{AbadL95Conjoin}. We prove it now in our own setting. It is Theorem~\ref{thm:grt-impl} below, but we need some considerations first.

For the same reasons that we avoid the \emph{next} temporal operator, we prefer to avoid explicit references to steps. For, if we later refine that \emph{next} step into a sequence of them, the reference to state $i+1$ turns out to be a moving one. The important concept here is that the partial path up to the present time is \emph{compatible} with the satisfaction of the formula, that is, that some maximal path that extends the partial one satisfies $\alpha$. We make this formal. The definition of when a path is a prefix (or initial segment) of another is the usual one, denoted by the symbol $<$, with reflexive closure $\le$. 

\begin{defn}[path compatible with formula]
Given $\calt\in\aets$, a finite path $\overline{g}$ in it, and a formula $\varphi$ in its language, we say that $\overline{g}$ is \emph{compatible} with $\varphi$, and denote it as $\overline{g}\compat\varphi$, iff there is some maximal path $\overline{g}'$ in $\calt$ such that $\overline{g}\le\overline{g}'$ and $\overline{g}'\models\varphi$.
\end{defn}
Given that $\calt$ and $\cala$ are going to be fixed, we spare them when writing $\overline{g}\compat\varphi$. Also, sometimes we write just $\overline{g}\models\varphi$ instead of the full $\calta,\overline{g}\models\varphi$.

According to this definition, if $\overline{g}$ is maximal, then $\overline{g}\compat\varphi$ iff $\overline{g}\models\varphi$.

For a simple example, consider a finite path $g_0\dots g_k$ such that at each $g_i$ the value of a certain Boolean property $p$ is \lstinline!true!. And suppose there are two possible ways that path can be maximally extended: $g_0\dots g_k g_{k+1}\dots$ and $g_0\dots g_k g'_{k+1}\dots$, with all unprimed stages still assigning \lstinline!true! to $p$, but $g'_{k+1}$ assigning \lstinline!false!. Then
\[g_0\dots g_k g_{k+1}\dots\models\tbox(p=\intt{true}),\]
and, therefore,
\[g_0\dots g_k\compat\tbox(p=\intt{true}),\]
even though
\[g_0\dots g_k g'_{k+1}\dots\not\models\tbox(p=\intt{true}).\]

Our definition of A/G satisfaction is somewhat involved, so we first give an intuitive explanation. Very informally, $\calta\models\alpha\grt\gamma$ is a promise from $\calt$ of not being the first to fail to perform its duties---if we see $\gamma$ as its duties and $\alpha$ as the environment's. Consider this diagram, showing two paths running from left to right, starting at the initial stage $g_0$.
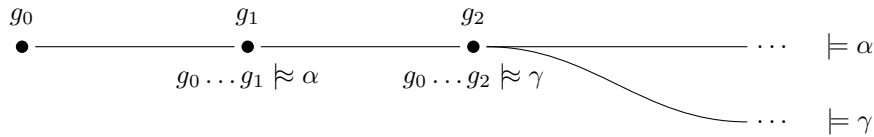
\begin{center}
\begin{tikzpicture}[auto]
\node [draw, circle, fill, inner sep=0, minimum size=1.5mm, outer sep=1mm] (g0) at (0, 0) {};
\node [draw, circle, fill, inner sep=0, minimum size=1.5mm, outer sep=1mm] (g) at (3, 0) {};
\node [draw, circle, fill, inner sep=0, minimum size=1.5mm, outer sep=1mm] (g'') at (6, 0) {};
\node (g') at (10, 0) {$\dots$};
\node (h) at (10, -1) {$\dots$};

\draw (g0) edge (g);
\draw (g) edge (g'');
\draw (g'') edge (g');
\draw [out=0, in=180] (g'') edge (h);

\node (xg0) at (0, 0.4) {$g_0$};
\node (xg1) at (3, 0.4) {$g_1$};
\node (xg) at (3, -0.4) {$g_0\dots g_1\compat\alpha$};
\node (xg2) at (6, 0.4) {$g_2$};
\node (xg'') at (6, -0.4) {$g_0\dots g_2\compat\gamma$};
\node (xg') at (11, 0) {$\models\alpha$};
\node (xh) at (11, -1) {$\models\gamma$};
\end{tikzpicture}
\end{center}

At present, $\calt$ and its environment have traversed together the path from $g_0$ to $g_1$, and have done so in a way compatible with the satisfaction of $\alpha$. The component $\calt$ cannot know what the environment is going to do in the future; it may choose to go along the upper path, that is, to keep on being compatible with $\alpha$. To ensure $\calta\models\alpha\grt\gamma$, the component $\calt$ has to keep on being compatible with $\gamma$ at least a little longer than the environment, for instance, until $g_2$.

In principle, the path that satisfies $\gamma$ needs not be the same one that satisfies $\alpha$, as shown in the diagram above. However, compatibility has to be preserved up to stages $g_1$ arbitrarily far in the future. The result is that the two branches get zipped into one.

\begin{defn}[path allowed in a compound]
A path $\overline{g}$ in $\calt\in\aets$ is said to be \emph{allowed} in $\calt\|_Y\cale$, for $\cale\in\ets$, if $\overline{g}$ is an element of some set of compatible paths in $\calt\|_Y\cale$.

In a similar way, a path $\overline{q}$ in $\calt\in\pts$ is said to be allowed in $\calt\|_Y\cale$, for $\cale\in\pts$, if there is a path in $\calt\|_Y\cale$ whose projection on $\calt$ is $\overline{q}$.
\end{defn}

\begin{defn}[A/G satisfaction]
\label{def:grt}
\begin{itemize}
\item
   For $\calt\in\aets$, an algebra $\cala$, and two formulas $\alpha$, $\gamma$, we define $\calta\models\alpha\grt\gamma$ by: for each egalitarian transition structure $\cale$ (the environment) and suitable $Y$, and for each finite path $\overline{g}$ in $\calt$ allowed in $\calt\|_Y\cale$ such that $\overline{g}\compat\alpha$, we have that:
   \begin{itemize}
   \item
      either $\overline{g}$ is maximal (hence $\overline{g}\models\alpha$) and then $\overline{g}\models\gamma$;
   \item
      or $\overline{g}$ is not maximal and, then, for each maximal $\overline{g}'$ with $\overline{g}<\overline{g}'$ and $\overline{g}'\models\alpha$, there is $\overline{g}''$ with $\overline{g}<\overline{g}''\le\overline{g}'$ and $\overline{g}''\compat\gamma$, that is, along each maximal extension that satisfies $\alpha$ (of which there must be some, because $\overline{g}\compat\alpha$) there is an intermediate path compatible with $\gamma$.
   \end{itemize}

\item
   For $\calt\in\pts$, the definition is, as usual, very similar to the above.

\item
   For $\calt\in\ets$, we define $\calta\models\varphi$ as equivalent to $\splitop(\calt),\cala\models\varphi$.
\item
   For rewrite systems of any kind, the definition is based on the transition structures which are their semantics, as usual.
\end{itemize}
\end{defn}

Possibly the simplest alternative concept of A/G satisfaction is $\calta\models\alpha\limpl\gamma$ (being $\limpl$ classical implication); that is, each path that satisfies the assumption also satisfies the guarantee. Though our definition of satisfaction is much more complex than that, \emph{a posteriori} both concepts turn out to be equivalent.

\begin{thm}[equivalence of $\grt$ and $\rightarrow$]
\label{thm:grt-impl}
In the conditions of Definition~\ref{def:grt}, \[\calta\models\alpha\grt\gamma \quad\text{iff}\quad \calta\models\alpha\limpl\gamma\]
and
\[\calr\models\alpha\grt\gamma \quad\text{iff}\quad \calr\models\alpha\limpl\gamma.\]
\end{thm}

\begin{proof}
First, we prove the theorem for $\calt\in\aets$, that is, for an atomic $\calt$. Assume $\calta\models\alpha\grt\gamma$, and let us prove that $\calta\models\alpha\limpl\gamma$. We have to show that each path $\overline{g}$ in $\calt$ which is maximal in $\calt$ and satisfies $\alpha$ also satisfies $\gamma$. We place $\calt$ in an arbitrary environment $\cale$ with empty synchronization criteria: $\calt\|_\emptyset\cale$. Certainly, $\overline{g}$ is allowed in $\calt\|_\emptyset\cale$ and is maximal in $\calt\|_\emptyset\cale$, because it is in $\calt$. Then, because $\calta\models\alpha\grt\gamma$ and the definition of A/G satisfaction, we have $\overline{g}\models\gamma$, as we wanted.

Now, assume $\calta\models\alpha\limpl\gamma$. Fix $\cale$ and $Z$. Fix also a path $\overline{g}$ in $\calt$ which is allowed in $\calt\|_Z\cale$ and is compatible with $\alpha$: $\overline{g}\compat\alpha$. If $\overline{g}$ happens to be maximal in $\calt$, then $\overline{g}\models\alpha$ and, because of the assumption, $\overline{g}\models\gamma$. Otherwise, if $\overline{g}$ if not maximal in $\calt$, fix a path $\overline{g}'$ which is maximal in $\calt$, extends $\overline{g}$ and satisfies $\alpha$. Because of the assumption, $\overline{g}'\models\gamma$. We can take $\overline{g}''=\overline{g}'$, and this completes the proof for atomic transition structures.

The same proof is almost verbatim valid for plain ones. And because satisfaction for composed structures is equivalent to the one for their splits, the result also holds for \pts. Finally, because satisfaction for rewrite systems is defined based on their semantics, the result also holds for the three types of rewrite systems.
\end{proof}

This is a most welcome result, because it means that we can use standard verification tools to perform compositional verification.

The particular case when $\alpha\equiv\intt{true}$ is worth stating.
\begin{corol}[true assumption]
In the conditions of Definition~\ref{def:grt},
\[\calta\models\mathtt{true}\grt\gamma \quad\text{iff}\quad \calta\models\gamma\]
and
\[\calr\models\mathtt{true}\grt\gamma \quad\text{iff}\quad \calr\models\gamma.\]
\end{corol}

After having Theorem~\ref{thm:grt-impl}, in view of our convoluted definition for A/G satisfaction, and considering also how trivial some of our examples are (maybe specially the one on chained buffers in Section~\ref{sec:buffers-cont} below), it is legitimate to ask if it would not have been better to use just implication to characterize A/G to begin with. The answer, in our opinion, is \emph{no}. The assertion $\cals\models\alpha\to\gamma$ is about the internal behavior of $\cals$; in contrast, $\cals\models\alpha\rhd\gamma$ is an assertion about $\cals$'s interaction with other systems. Their equivalence (in appropriate conditions) is a fortunate, \emph{a posteriori} fact. Perhaps it could be likened to the equivalence between $\vdash\varphi\to\psi$ and $\varphi\vdash\psi$ in classical first-order logic.

We finish this section with the theorem which justifies the soundness of A/G.

\begin{thm}[soundness of A/G]
\label{thm:ded-rule}
With the notational conventions used so far, let $\calr_n$ ($n=1,\dots,N$) be rewrite systems of any of the kinds discussed in this work, and let $\calr$ be their composition with respect to the synchronization criteria $Y$, $\calr=\|_Y\calr_n$. If all the following hold:
\begin{enumerate}
\setlength\itemsep{2pt}
\item for each $n=1,\dots,N$ and each $i=1,\dots,\ell_n$, we have that
    \begin{enumerate}
    \item\label{item:component-ag}
       $\calr_n\models\alpha_{ni}\grt\gamma_{ni}$,
    \item\label{item:long-enough}
       $\calr_n\models\alpha_{ni}\limpl\gamma_{ni}$ implies $\calr\models\alpha_{ni}\limpl\gamma_{ni}$,
    \end{enumerate}
\item\label{item:proviso} $\left(\Land_{n=1}^N\Land_{i=1}^{\ell_n}\alpha_{ni}\limpl\gamma_{ni} \text{ and } \Land_{(p,p')\in Y} p=p'\right)$ imply $\alpha\limpl\gamma$,
\end{enumerate}
then $\calr\models\alpha\grt\gamma$.
\end{thm}

\begin{proof}
Because of Theorem~\ref{thm:grt-impl}, for each $i$, we have $\calr_1\models\alpha_{1i}\grt\gamma_{1i}$ iff $\calr_1\models\alpha_{1i}\limpl\gamma_{1i}$ and, because of Condition~\ref{item:long-enough}, this implies $\|_Y\calr_n\models\alpha_{1i}\limpl\gamma_{1i}$. The same reasoning holds for the other components $\calr_n$ and its A/G statements in view of Condition~\ref{item:component-ag}. Additionally, $\|_Y\calr_n\models p_1=p_2$ for each $(p_1,p_2)\in Y$, because of the very definition of the synchronous composition and of satisfaction. Thus, $\|_Y\calr_n$ satisfies all conjuncts in the left-hand side of Condition~\ref{item:proviso}. And, thus, it satisfies the right-hand side, that is, $\|_Y\calr_n\models\alpha\limpl\gamma$ which, again because of Theorem~\ref{thm:grt-impl}, is equivalent to $\|_Y\calr_n\models\alpha\grt\gamma$.
\end{proof}

Each of the conditions included in Condition~\ref{item:component-ag} asks for an A/G statement to hold in a component. Often, a single A/G statement is asked from each component, that is, $\ell_n=1$ for some or all $n$. In particular, the statement that $\calr_1\models\varphi$ implies $\calr_1\|_Y\cale\models\varphi$, can be seen as the particular case where $n=2$, $\ell_1=1$, $\ell_2=0$, $\calr_2=\cale$, and $\alpha_{11}=\alpha=\intt{true}$.

This theorem allows to reduce the proof of an A/G statement on a composed system to similar proofs on smaller systems, plus checking the validity of an LTL formula. The word \emph{reduce} in the previous sentence is questionable, because the number of tasks seems to have increased. In addition, obtaining the formulas needed in the premises is not always easy. (We will have something more to say about this in Section~\ref{sec:concl}.) The positive side is that each statement has to be proved now in a smaller model. And that, once proved, each component can be reused with no need for new proofs.

We want to remark that Maude provides the tools needed for compositional verification using Theorem~\ref{thm:ded-rule}: the model checker can be used to verify $\calr_1\models\alpha_{11}\limpl\gamma_{11}$ and the other similar results, and the tautology checker can be used to check the validity of the final formula. Maude is not able to handle our properties, so the formulas must be transformed to use only Boolean propositions as discussed in Section~\ref{sec:back-standard}.

Condition~\ref{item:long-enough} holds for safety formulas, as shown in Proposition~\ref{prop:safety}, or in the presence of fairness between components and absence of emerging deadlocks, as shown in Proposition~\ref{prop:long-enough}. A case of interest is when the assumption $\alpha$ implies such fairness and deadlock freeness requirements. An instance of this is the following. In each $\calr_n$, there is a property $t_n$ defined so that it holds true at each transition and false at each state of $\calr_n$. In addition, the formula $\alpha$ includes as a conjunct (or implies otherwise) the formula $\varphi_n=\Land_n(\tbox\tdiam t_n=\intt{true} \land \tbox\tdiam t_n=\intt{false})$, which means that component $\calr_n$ advances infinitely often, which implies by itself fairness between components and absence of emerging deadlocks. Those formulas $\varphi_n$ need not be appropriate for every case. With our definitions, terminating systems may be fair and still not satisfy $\varphi_n$. In each case, more refined formulas may be better suited.

It may be worth noting, to avoid confusion, that $\calr_n\models\varphi_n$ for all $n$ does not entail fairness or deadlock freeness. The reason is that it may happen that $\calr_n\models\varphi_n$ but $\calr\not\models\varphi_n$ because, well, deadlocks or lack of fairness. Things are different when we use the $\varphi_n$, not as guarantees, but as assumptions, which we did in the previous paragraph.

There is another way to verify a compositional specification, which is to split it into a monolithic one and use standard verification techniques on it. This works thanks to the following proposition.

\begin{prop}
\label{prop:mono-verif}
With the notational conventions used so far, for $\calr_n$ egalitarian rewrite systems for $n=1,\dots,N$, we have that
\[\|_Y\calr_n\models\alpha\grt\gamma \quad\text{iff}\quad \splitop(\|_Y\calr_n)\models\alpha\grt\gamma.\]
\end{prop}

\begin{proof}
This is an easy corollary of Theorem~\ref{thm:grt-impl} and Proposition~\ref{prop:split-sat}.
\end{proof}

\subsection{Example: chained buffers (continued)}
\label{sec:buffers-cont}

This continues the example from Section~\ref{sec:buffers}. It is immediate to prove that each of the buffers satisfies $\intt{BUFFER}n\models\tdiam\intt{isReceiving}\to\tdiam\intt{isSending}$. Therefore, by Theorem~\ref{thm:grt-impl}, $\intt{BUFFER}n\models\tdiam\intt{isReceiving}\grt\tdiam\intt{isSending}$. We expect to be able to prove a similar behavior for the whole chain of buffers, \lstinline!3BUFFERS!, using Theorem~\ref{thm:ded-rule}. Concretely, in this case:
\begin{itemize}
\item
   $N=3$,
\item
   $\calr_n=\intt{BUFFER}n$,
\item 
   $\ell_n=1$,
\item
   $\alpha_{n1} = \tdiam\intt{BUFFER}n\intt{\textcolor{red!50!black}{\$}isReceiving}$, $\gamma_{n1} = \tdiam\intt{BUFFER}n\intt{\textcolor{red!50!black}{\$}isSending}$,
\item
   $\alpha = \tdiam\intt{BUFFER1}\intt{\textcolor{red!50!black}{\$}isReceiving}$, $\gamma = \tdiam\intt{BUFFER3}\intt{\textcolor{red!50!black}{\$}isSending}$.
\end{itemize} 
Regarding the conditions in Theorem~\ref{thm:ded-rule}: Condition~\ref{item:component-ag} (that is, $\calr_n\models\alpha_{n1}\grt\gamma_{n1}$ for each $n$) has already been justified, and Condition~\ref{item:proviso} (implication for temporal formulas) is easily seen to hold. Fairness is ensured by the way the buffers synchronize, and we hope it is clear that no deadlocks can emerge, so also Condition~\ref{item:long-enough} holds. From which we can deduce
\[\intt{3BUFFERS} \models \tdiam\intt{BUFFER1}\intt{\textcolor{red!50!black}{\$}isReceiving} \grt \tdiam\intt{BUFFER3}\intt{\textcolor{red!50!black}{\$}isSending}.\]
The properties \lstinline!BUFFER1$isReceiving! and \lstinline!BUFFER3$isSending! can be better seen here as properties of \lstinline!3BUFFERS!, and our extension to Maude's syntax allows to define synonyms, so that the above can also be written as
\[\intt{3BUFFERS} \models \tdiam\intt{3BUFFERS}\intt{\textcolor{red!50!black}{\$}isReceiving} \grt \tdiam\intt{3BUFFERS}\intt{\textcolor{red!50!black}{\$}isSending}.\]

\subsection{Example: mutual exclusion (continued)}
\label{sec:mutex-cont2}

We finish now our discussion of the example from Sections~\ref{sec:mutex} and~\ref{sec:mutex-cont}. It is easy to prove, either by model checking or by mere inspection, that
\[\intt{MUTEX}\models\tbox\neg(\intt{isGranting(1)}\land\intt{isGranting(2)}).\]
Reasoning intuitively, we know that the same formula holds when \lstinline!MUTEX! is made a component of \lstinline!SAFE-TRAINS!. And, because the composition requires each \lstinline!isGranting! property to be synchronized with the corresponding \lstinline!isCrossing!, we deduce
\begin{equation*}
\intt{SAFE-TRAINS}\models\tbox\neg(\intt{S-TRAIN1\textcolor{red!50!black}{\$}isCrossing}\land\intt{S-TRAIN2\textcolor{red!50!black}{\$}isCrossing}).
\end{equation*}

Formally, we have used Theorem~\ref{thm:ded-rule} with
\begin{itemize}
\item 
   $N=3$,
\item
   $\calr_1=\intt{S-TRAIN1}$, $\calr_2=\intt{S-TRAIN2}$, $\calr_3=\intt{MUTEX}$,
\item
  $\ell_1=\ell_2=0$, $\ell_3=1$,
\item 
  $\alpha_{11}=\intt{true}$, $\gamma_{11}=\tbox\neg(\intt{isGranting(1)}\land\intt{isGranting(2)})$,
\item 
  $\alpha=\intt{true}$, $\gamma=\tbox\neg(\intt{S-TRAIN1\textcolor{red!50!black}{\$}isCrossing}\land\intt{S-TRAIN2\textcolor{red!50!black}{\$}isCrossing})$.
\end{itemize}

Component fairness does not always hold, but the formulas involved are safety ones, which is enough according to Proposition~\ref{prop:safety}.

It was observed before that the system \lstinline!MUTEX! can be seen as a controller or strategy. The verification task is, in this case, of a different nature from the one in the previous example on chained buffers, in which the behavior of the compound necessarily results from the interactions between the components.

\subsection{Example: crossing the river (continued)}
\label{sec:river-cont}

We now verify the example whose compositional specification was given in Section~\ref{sec:river}. 

We want to prove that the composed system satisfies the formula $\tdiam\intt{success}$, where the property is defined in \lstinline!RIVER! like this:
\begin{lstlisting}
ppt success : -> Bool .
eq success @ (noBelong |~| mark farmer wolf goat cabbage) = true .
eq success @ G = false [owise] .
\end{lstlisting}
We consider a \lstinline!success! that all items, including the \lstinline!mark!, are on the same side of the river. So, our hypothesis is that each possible sequence of crossings executed according to our two guidelines leads to a valid solution.

Our implementation allows to transform the four-component system into a single standard one (that is, to apply the split), and use Maude's model checker in the resulting system. This split approach has the advantage that we do not need to find the A/G statements for each component. However, in this case the needed A/G statements for \lstinline!RIVER-W-PREV!, \lstinline!AVOID1!, and \lstinline!AVOID2! are quite clear. So, we work compositionally on those three, although we use the split below to verify the two-component system \lstinline!RIVER-W-PREV!. Namely, we need \lstinline!RIVER-W-PREV! to satisfy the A/G statement \[\intt{RIVER-W-PREV}\models\tbox\neg\intt{danger}\land\tbox\neg\intt{undoing} \;\grt\; \tdiam\intt{success}.\]
In words: success is eventually reached assuming the environment allows neither dangerous situations nor undoings. If we are able to prove this, and having into account that, quite obviously, \lstinline!AVOID1! and \lstinline!AVOID2! satisfy $\tbox\neg\intt{avoid}$, we can use Theorem~\ref{thm:ded-rule} to deduce that the four-component system satisfies $\tdiam\intt{success}$. Thus, we perform the split on \lstinline!RIVER-W-PREV! and model check it for $\tbox\neg\intt{danger}\land\tbox\neg\intt{undoing} \;\rightarrow\; \tdiam\intt{success}$, as allowed by Theorem~\ref{thm:grt-impl}.

The result is that the formula \emph{does not hold}, and the model checker hands us a counterexample: an infinite execution that never gets to the desired state. On inspection, we find out that the problem with our solution stems from a symmetry between the roles of the wolf and the cabbage. For example, suppose we are in this situation (in which we omit the \lstinline!mark!, because its location does not make any difference):
\begin{lstlisting}
farmer wolf goat |~| cabbage
\end{lstlisting}
Then, the farmer crosses with the wolf, to get
\begin{lstlisting}
goat |~| farmer wolf cabbage
\end{lstlisting}
and, then, the farmer crosses back with the cabbage to get \begin{lstlisting}
farmer cabbage goat |~| wolf
\end{lstlisting}
The new situation is symmetric to the first one, because the roles of cabbage and wolf are similar: eating can take place whenever the goat is left unattended with any of them. As critical as the difference may be for the goat itself, it is irrelevant for us who eats whom. Indeed, if a solution is obtained for a specific situation, the corresponding symmetric solution can be applied to the symmetric situation.

At the end, what we need is to strengthen the concept of undoing to avoid also symmetric movements, which we get by adding two equations to the definition of the property \lstinline!undoing!:
\begin{lstlisting}
eq undoing @ (wolf > cabbage) = true .
eq undoing @ (cabbage > wolf) = true .
\end{lstlisting}
Now, the A/G satisfaction holds, showing that the strengthened guidelines are sufficient. Indeed, only two solutions are left, one symmetric to the other, both optimal in their number of moves. And, fixed this part of the composed system, we already know the whole compound works.

\section{Additional examples}
\label{sec:add-ex}

The examples used in this paper up to now have been chosen to be illustrative, so they are rather simple on purpose. Because of such simplicity, we have been able to omit many details of our implementation, which are unimportant for the theoretical work here presented. In this section, we offer a cursory overview of two more complex examples which were presented and discussed at length in~\cite[Chapter~7]{Mart21Thesis}. Our presentation in this section is necessarily incomplete. The source code for the examples is available online~\cite{Mart20Web}. Both examples have been run through our prototype implementation and the results have been verified using Maude's toolset with the techniques described in this paper.

It has been mentioned that there are two ways to verify a compositional rewriting-logic specification. The first is to perform a compositional verification, according to Theorem~\ref{thm:ded-rule}, using A/G assertions for each component. This is the approach taken in the examples shown so far in this paper, where the A/G assertions for the components were easily found. The second way, justified by Proposition~\ref{prop:mono-verif}, consists in splitting the compositional specification to obtain an equivalent monolithic one which is then monolithically verified. This is the technique illustrated in the new examples we are presenting next. In them, finding the A/G assertions for the component systems turns out to be a difficult task (mainly because there is no general controller, but rather an emergent behavior). So we have used our prototype implementation of the split to obtain a standard Maude module, which is the one we have actually verified using Maude's toolset.

Questions about performance are analyzed in~\cite[Chapter~7]{Mart21Thesis}. We quote: ``our implementation needs more than two minutes to process the ABP specification and produces more than 18,000 rules, while the Needham-Schroeder example is processed in only two seconds and produces less than 600 rules.'' These numbers are largely dependent on the implementation we are using. Again, more details are in~\cite{Mart21Thesis}.

\subsection{Alternating bit protocol}
\label{sec:abp}

The first example is a specification of the alternating bit protocol, ABP from now on, to send messages reliably on a channel which may lose some of the packets it receives. We consider an ABP system as consisting of four interacting components:
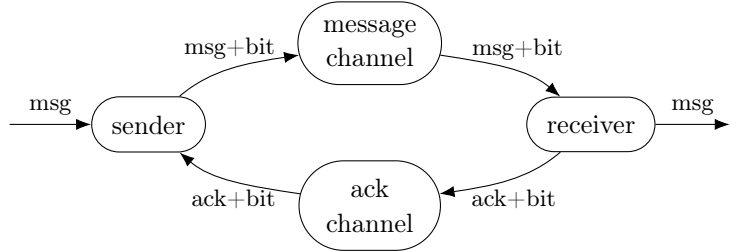
\begin{center}
\begin{tikzpicture}[auto, xscale=0.42, yscale=0.27, every text node part/.style={align=center}]]
\node [inner sep=7pt] (prdr) at (2, 0) {};
\node [state, inner sep=7pt] (sndr) at (7, 0) {sender};
\node [state, inner sep=7pt] (msgchnl) at (14, 4) {message\\channel};
\node [state, inner sep=7pt] (ackchnl) at (14, -4) {ack\\channel};
\node [state, inner sep=7pt] (rcvr) at (21, 0) {receiver};
\node [inner sep=7pt] (csmr) at (26, 0) {};

\draw [-{Latex[length=2mm]}] (prdr) edge node {\small msg} (sndr);
\draw [-{Latex[length=2mm]}, bend left=15, out=25] (sndr) edge node [xshift=7mm] {\small msg+bit} (msgchnl);
\draw [-{Latex[length=2mm]}, bend left=15, in=180-25] (msgchnl) edge node [xshift=-6mm] {\small msg+bit} (rcvr);
\draw [-{Latex[length=2mm]}, bend left=15, out=25] (rcvr) edge node [xshift=-6mm] {\small ack+bit} (ackchnl);
\draw [-{Latex[length=2mm]}, bend left=15, in=180-25] (ackchnl) edge node [xshift=7mm] {\small ack+bit} (sndr);
\draw [-{Latex[length=2mm]}] (rcvr) edge node {\small msg} (csmr);
\end{tikzpicture}
\end{center}
The sender and the receiver are the components which implement the protocol. There are two channels: one for transmitting messages it gets from the sender; the other for transmitting acknowledgments back. The internal workings of the two channels are the same. Missing, at the two ends of the diagram, are a producer and a consumer. Thus, the result of our four-component specification is meant to be used in turn as a component in a larger system.

When specifying systems of some complexity, we like to enforce modularity further by using the syntax for parametric specification in Maude~\cite{MartVM2018ParamCompSpec}. In our case, the final module, which represents the whole ABP system, is specified as
\begin{lstlisting}
emod ABP-BP{Sndr    :: SENDER-IF,
            MsgChnl :: CHANNEL-IF,
            AckChnl :: CHANNEL-IF,
            Rcvr    :: RECEIVER-IF} is
   sync Sndr || MsgChnl || AckChnl || Rcvr
      on ...
endem
\end{lstlisting}
We have hidden the synchronization criteria. The important point here is that the module \lstinline!ABP-BP! is parametric in the four modules it receives, each modeling one of the components: a sender, two channels, and a receiver. The suffix \lstinline!-BP! stands for \emph{blueprint}, and \lstinline!-IF! stands for \emph{interface}. For example, \lstinline!Sndr! is the name given to the first formal parameter, which is to be instantiated with a module that implements the interface specified in \lstinline!SENDER-IF!. Namely, \lstinline!SENDER-IF! (which is called a \emph{theory} in Maude jargon) includes the declaration of the following five properties:
\begin{lstlisting}
ppt msgIn : -> Msg? .
ppts msgPckOut ackPckIn : -> Packet? .
ppts canAckChnlPass canMsgChnlGet : -> Bool .
\end{lstlisting}
The interfaces contain declarations, and no implementation. Then, whichever module defines these properties is valid as the first argument to build an ABP. In the end, after we have specified the needed modules which fit the interfaces, with respective module names \lstinline!Sender!, \lstinline!MsgChannel!, \lstinline!AckChannel!, \lstinline!Receiver!, we obtain the final result with
\begin{lstlisting}
emod ABP is
   inc ABP-BP{Sender, MsgChannel, AckChannel, Receiver} .
endem
\end{lstlisting}

For the two channels, their specification includes the possibility that messages are lost. As a consequence, a fairness constraint is required for each channel, to ensure that at least some messages get through. Namely, we use the assumption $\tbox\tdiam\intt{isPassing}$, where \lstinline!isPassing! is a Boolean property defined to be true exactly when a message is going out of the channel. We also need fairness assumptions on the sender and the receiver, which we do not care to show here.

For verification, given those assumptions, we want to prove the following formula:
\[\gamma\;=\;
\tbox ( \intt{isAccepting}
        \limpl
        ( \intt{isAccepting}
          \tu
          ( \neg\,\intt{isAccepting} \tu \intt{isDelivering} ) ) ) .
\]
The property \lstinline!isAccepting! is meant to be true whenever the sender gets a message from some producer process (that is, when the sender is executing a transition to the purpose of getting such a message); similarly, \lstinline!isDelivering! is true whenever the receiver gives the message to some consumer process. In words, this says that to each input to the ABP system follows an output, with no other input in between. This can also be interpreted as saying that ``at the next stage of interest'' \lstinline!isDelivering! holds (see discussion in Section~\ref{sec:ltl}).

If we call $\alpha$ the conjunction of the four fairness constraints mentioned above, we want to prove $\intt{ABP}\models\alpha\grt\gamma$. To prove it, we first use our prototype implementation to obtain a monolithic standard Maude module equivalent to the original compositional specification of \lstinline!ABP!, then we verify on the resulting module the formula $\alpha\to\gamma$ using Maude's model checker.

\subsection{Needham-Schroeder public-key protocol}

Needham-Schroeder is a public-key protocol for safe communication between two actors: an initiator and a responder. It is known to be unsafe in the presence of an attacker, but we are here interested in the simple case where there is no attacker.

This example illustrates the convenience of compositionality in system specification in two ways. First, the two actors (initiator and responder, Alice and Bob) are specified as independent systems, and only later made to interact. Second, and more to the point of the purpose of this example, each actor is specified as a base module describing all its nondeterministic capabilities (sending, encrypting\dots), which is then controlled by another module making it behave actually as initiator or responder. The technique we use to make possible this control is, roughly speaking, the use of a language by which the controller sends commands to the base module. This is achieved, of course, through the use of properties and synchronous composition.

This diagram shows the components we model in Maude:
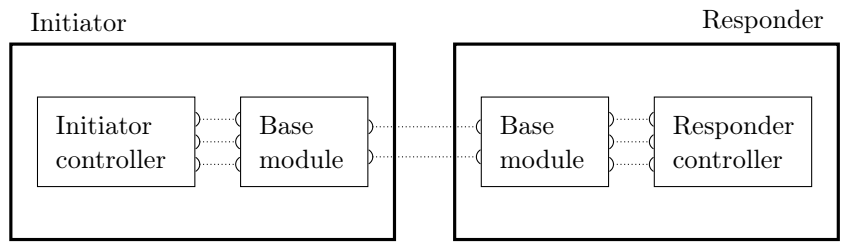
\begin{center}
\begin{tikzpicture}[auto]
\node [draw, inner sep=7pt, minimum height=1cm, text width=16mm] (init) at (0, 0) {Initiator controller};
\node [draw, inner sep=7pt, minimum height=1cm, text width=12mm] (base1) at (2.5, 0) {Base module};
\node [draw, inner sep=7pt, minimum height=1cm, text width=12mm] (base2) at (5.7, 0) {Base module};
\node [draw, inner sep=7pt, minimum height=1cm, text width=16mm] (resp) at (8.2, 0) {Responder controller};

\draw ([yshift=3mm]init.east) edge [)-(, out=0, in=180, densely dotted] ([yshift=3mm]base1.west);
\draw ([yshift=0mm]init.east) edge [)-(, out=0, in=180, densely dotted] ([yshift=0mm]base1.west);
\draw ([yshift=-3mm]init.east) edge [)-(, out=0, in=180, densely dotted] ([yshift=-3mm]base1.west);

\draw ([yshift=3mm]base2.east) edge [)-(, out=0, in=180, densely dotted] ([yshift=3mm]resp.west);
\draw ([yshift=0mm]base2.east) edge [)-(, out=0, in=180, densely dotted] ([yshift=0mm]resp.west);
\draw ([yshift=-3mm]base2.east) edge [)-(, out=0, in=180, densely dotted] ([yshift=-3mm]resp.west);

\draw ([yshift=2mm]base1.east) edge [)-(, out=0, in=180, densely dotted] ([yshift=2mm]base2.west);
\draw ([yshift=-2mm]base1.east) edge [)-(, out=0, in=180, densely dotted] ([yshift=-2mm]base2.west);

\draw [very thick] (-1.4, -1.3) rectangle ++(5.1, 2.6);
\node (kk1) at (-0.5, 1.6) {Initiator};

\draw [very thick] (4.5, -1.3) rectangle ++(5.1, 2.6);
\node (kk2) at (8.6, 1.6) {Responder};
\end{tikzpicture}
\end{center}
Each of the small arcs represents a property, and the dotted lines represent  synchronization. The initiator and the responder both implement the interface theory called \lstinline!ACTOR-IF!. The base modules, which we call \lstinline!INITIATOR-BASE! and \lstinline!RESPONDER-BASE!, are exactly the same except for their initial states. The blueprint for the total system is this:
\begin{lstlisting}
emod NSPKP-BP{I :: ACTOR-IF, R :: ACTOR-IF} is
   sync I || R
      on R$msgRcv := I$msgSnd
      /\ I$msgRcv := R$msgSnd .
   ...
   ag True |> [] <> isCommEstablishedInI /\ [] <> isCommEstablishedInR .
endem
\end{lstlisting}
This is again a parameterized module, which has to be fed with implementations for the initiator \lstinline!I! and the responder \lstinline!R!. The sentence with the \lstinline!ag! keyword is an A/G assertion saying that, with no assumption (\lstinline!True!), the module has to guarantee that communication is established arbitrarily often for both actors. (The ABP example also had A/G assertions, which we preferred to omit to simplify the presentation.) The symbol \lstinline!:=! in the synchronization criteria is, for our purposes here, equivalent to the \lstinline!=! we have used all the time.

To implement control, we have used a very simple language of commands that the controller issues and the base module executes. For example, the \lstinline!INITIATOR! is specified as the composition of \lstinline!INITIATOR-PROTOCOL! and \lstinline!INITIATOR-BASE!:
\begin{lstlisting}
emod INITIATOR is
   sync INITIATOR-PROTOCOL || INITIATOR-BASE
      on INITIATOR-BASE$action := INITIATOR-PROTOCOL$action
      /\ INITIATOR-BASE$arg := INITIATOR-PROTOCOL$arg
      /\ INITIATOR-PROTOCOL$isErrorState := INITIATOR-BASE$isErrorState .
   ...
endem
\end{lstlisting}
Thus, the base module \lstinline!INITIATOR-BASE! receives from the controller (by synchronizing properties) the \lstinline!action! to be performed and the \lstinline!arg!uments on which to perform them. The feedback to the controller is whether there has been some error (namely, an unsuccessful decryption). The states of the base system are given by a set of pairs key-value. For example, this is the initial state for the \lstinline!INITIATOR!:
\begin{lstlisting}
eq init = ('myid : alice) ('xid : bob)
          ('myprivkey : priv(alice)) ('xpubkey : pub(bob))
          ('mynonce : nonce(alice)) .
\end{lstlisting}
Thus, the key \lstinline!'myid! is storing the value \lstinline!alice!, and so on. There are rules in the base module to specify the different actions it is able to perform: send, receive, decrypt, encrypt, and check whether the values stored under two given indices are equal. For example:
\begin{lstlisting}
rl D =[ sending | D ]=> D .
rl D =[ receiving(M) | D ]=> D <+ ('msg : M) .
\end{lstlisting}
So, sending leads to no change in the values stored, but receiving adds or overwrites a pair with the key \lstinline!'msg! and the value of whatever it received. Slightly more complex rules implement the rest of the capabilities.

Then, these are the rules for the \lstinline!INITIATOR-PROTOCOL! that makes \lstinline!INITIATOR-BASE! behave as an actual initiator for the protocol:
\begin{lstlisting}
rl 1 =[ 1 : encrypt('mynonce 'myid)         ]=> 2 .
rl 2 =[ 2 : send                            ]=> 3 .
rl 3 =[ 3 : receive                         ]=> 4 .
rl 4 =[ 4 : decrypt('recmynonce 'recxnonce) ]=> 5 .
rl 4 =[ 4 : decrypt('recmynonce 'recxnonce) ]=> error .
rl 5 =[ 5 : check('recmynonce, 'mynonce)    ]=> 6 .
rl 5 =[ 5 : check('recmynonce, 'mynonce)    ]=> error .
rl 6 =[ 6 : encrypt('recxnonce)             ]=> 7 .
rl 7 =[ 7 : send                            ]=> 8 .
rl 8 =[ 8 : reset                           ]=> 1 .
\end{lstlisting}
The states are represented by mere numbers. But the interesting part is that the initiator part of the protocol can be read line by line in the transition terms: first, encrypt the values stored under the keys \lstinline!'mynonce! and \lstinline!'myid!, then send the result of the encryption, and so on. The way the properties are defined and the way the synchronization is specified ensures that the \lstinline!send! in the controller is executed synchronized with the \lstinline!sending! in the base module. The wildly nondeterministic behavior of the base system is transformed into an almost fully deterministic one once synchronized with the controller.

For verification, as in the previous example, we transform the compositional specification into a monolithic, standard one and, then, use Maude's model checker to verify the formula given in the \lstinline!ag! statement. Again, this method is fast and simple and avoids the costly search for the components' A/G assertions.

\section{Closing material}
\label{sec:closing}

\subsection{Related work}
\label{sec:related}

We are not aware of any other work dealing precisely with compositional verification and rewriting logic, but certainly our work on compositionality, both for specification and for verification, is inspired by others, including process algebras, coordination models, and many more. Our results on A/G are also strongly based on existing work for other settings~\cite{ElkaG+18CircAG,JonsY96AGinLTL,AbadL95Conjoin}.

Besides A/G, many other verification techniques are discussed in the literature which are compositional in nature. Often they consist in simplifying the isolated components before composing them into a single global system. This is related to our work on simulation and abstraction in Section~\ref{sec:simul}. Simplification is performed either taking into account the behavior of the environment, or the temporal formula to be proved, or both. Sometimes, the global system is not even produced in full, but instead the global state-space is created on the fly traversing in parallel the components. The paper~\cite{GaravelLM2015CompVerifCADP} describes these and other techniques and their implementation in the toolset CADP~\cite{CADP2023Web}. On this same matter of simplifying a component before using it,~\cite{AndreKOP2012CountIncrVerif} proposes the use of Symbolic Observation Graphs, similar to the predicate abstractions we mentioned also in Section~\ref{sec:simul}. All these works use LTSs to model processes.

In the field of Petri nets,~\cite{KlaiHI2005ModVerifPN} deals with decomposing a Petri net into smaller ones. Isolated component nets, detached from the rest, are enriched with \emph{abstraction places} representing, in a sense, the environment. It also discusses \emph{non-constraining} interactions between components, a concept similar to our requirement of fairness and deadlock freeness. Their conclusion is worth quoting: ``experimental results show that this technique is efficient for some models, but for others the combinatorial explosion is not really attacked.'' Similar thoughts are expressed by~\cite{GaravelLM2015CompVerifCADP} and endorsed by our own experience.

\subsection{Future work}

The most substantial path we would like to explore in the future is the possibility of implementing strategies by synchronous composition. We see strategies in a broad sense, encompassing controllers, protocols, monitors, coordinators\dots\ Strategies are applied to nondeterministic systems to guide them, reducing or removing their nondeterminism. The rules of chess allow for many movements from each position. On that, a good strategy reduces the possibilities to probably just one at each point in a match. In the same way, when specifying the behavior of systems, we can specify a base system with all its nondeterministic capabilities and, then, use it under the control of a strategy; even in different ways under different strategies. This idea has been used with Maude and its strategy language to implement Knuth-Bendix-like completion as a basic set of correct rules on which different strategies are applied~\cite{Lesc89Complet,ClavM97IntStrat,VerdM11ComplStrat}, also for congruence closure~\cite{BachTV03CongrCl}, and for specifying insertion sort as a base system with a single rule for swapping cell contents which is then conveniently controlled~\cite{MartMV05StratLang,EkerM+07DedStratRw}. Our examples in this paper and in previous work~\cite{Mart21Thesis,MartVM20CompSpec} can be also viewed in this way.

Besides that, and being more concrete, we lack a proof that the procedure in Theorem~\ref{thm:ded-rule} is complete. That is, it is not proved whether, given $\alpha$ and $\gamma$, appropriate formulas $\alpha_{ni}$, $\gamma_{ni}$ can always be found. Based on similar results in similar contexts, we conjecture it is complete, but a proof is currently missing.

Adding other similar rules would also enrich our work. In particular, circular deduction rules~\cite{ElkaG+18CircAG} are different enough to deserve our attention. We mean, from $\calr_1\models\varphi\grt\varphi'$ and $\calr_2\models\varphi'\grt\varphi$, deduce $\calr_1\|\calr_2 \models \Phi$ for some formula $\Phi=\Phi(\varphi,\varphi')$.

Some works~\cite{CoblGP03LearnAss,BobaPG08AGAbstr} have shown how A/G reasoning can be automated. And also abstraction can be automated, for example, with the technique known as \emph{counter-example guided abstraction refinement} (CEGAR)~\cite{ClarG+00CEGAR,ChakC+04SEMC}. This can even be applied to a compositional system specification. Adding any such automation to our implementation would increase its usefulness. Specially the generation of intermediate formulas, because that would mean we have a new completely automated way to verify systems.

Our prototype implementation can be advanced in several ways to make it more complete, efficient, reliable, and easy to use. Also, translating compositional specifications in extended Maude to, for example, CADP syntax would allow the use of the rich CADP toolset for compositional verification.

\subsection{Conclusion}
\label{sec:concl}

There are reasons to be skeptic about the value of compositional verification, and in particular about the A/G technique. The main reason is the difficulty of finding the needed \emph{intermediate} formulas: in a simple case, a compositional proof of $\calr_1\|_Y\calr_2\models\varphi$ requires finding a formula $\gamma$ such that $\calr_1\models\gamma$ and $\calr_2\models\gamma\grt\varphi$. Finding such a formula $\gamma$, or whatever is needed in more complex cases, is in general a difficult task. We have chosen the examples in this paper so that those intermediate formulas are easily found. Examples in previous work, mentioned in Section~\ref{sec:add-ex}, showed other examples for which the intermediate formulas were not obvious at all, which made us prefer to use the split, finishing with a monolithic verification for a compositional specification.

Techniques have been devised for automatically generating such intermediate formulas~\cite{ElkaG+18CircAG,CoblGP03LearnAss,BobaPG08AGAbstr}. However, an experimental study~\cite{CoblAC06DecompAG} on the efficiency of these techniques with two actual tools draws this conclusion: ``This discouraging result, although preliminary, raises doubts about the usefulness of assume-guarantee reasoning.'' In a different style, a computation of the theoretical complexity of A/G~\cite{KupfV00AutModMC} finds it to be quite large: ``The results of this paper indicate that modular model checking [\dots] is rather intractable.'' Additionally, not many of the well-known tools for verification include the possibility of compositional verification, notable exceptions being BIP~\cite{BasuBS08BIP} and TLA$^+$~\cite{Lamp02SpecSys}. Whether this is because their practitioners have not found the need for it or for some other reason, we cannot say. To this, we can add our own, limited experience trying to perform compositional verification within our proposed framework, from which we have learned that the generation of temporal formulas for components is a laborious task. Moreover, the use of Theorem~\ref{thm:ded-rule}, in its Condition~\ref{item:proviso}, requires checking that a certain LTL formula, potentially large, is a tautology. We have used Maude's tautology checker, which in some cases takes very long to reach an answer. In all, if we want to verify a compositionally specified system, the cheaper way, both in human time and in computer time, may well be transforming it into a monolithic one (through the split operation), and performing monolithic model checking on the result. There is ongoing work in this area, so improvements can be expected. It may seem, however, that we need to justify our work on compositional verification. We devote a few lines to it.

First, we have already mentioned at the end of Section~\ref{sec:related} that both~\cite{KlaiHI2005ModVerifPN} and~\cite{GaravelLM2015CompVerifCADP} have found their compositional techniques (which do not include A/G) to be more effective than monolithic ones in some, though not all, cases.

One of our goals in this work was to show that compositional reasoning in rewriting logic is possible based on our framework for compositional specification. Componentwise abstraction and simulation and the A/G technique, in addition to whatever value they may have by themselves, were chosen by us as case studies to put our framework to the test. After having written this paper, we feel confident that new developments could be adapted as well.

The discouraging studies mentioned above miss a key ingredient of modularity, namely, reuse. They consider compositional verification as if it has to be completely redone from scratch every time. But, once the design of a system has been carried out modularly, the temporal formulas needed from each component have been determined, and the proof that some global formula follows from those of the components has been completed, all that is valid forever. If one of the components has to be modified, refined, or replaced, only the new component needs verification against the formulas already known to be needed from it.

A library of ready-to-use components is another instance of the convenience of modular design and verification. We have already remarked that we are specially interested in studying how strategies can be implemented as components that exert their control by means of synchronous composition. In these cases, the component implementing a strategy has to be independent of the rest, and has to perform its task whatever system it happens to be attached to. Thus, because the mutual exclusion controller in our example satisfies the mutual exclusion property, so does any composed system that relies on it. No need to find intermediate formulas or check complex provisos. In contrast, it is in cases when the global behavior is emergent, as in the ABP example in Section~\ref{sec:abp}, that finding the intermediate formulas is a difficult task.

The above discussion has to do with verification. In specification (or design, or modeling) the value of compositionality is less controversial.

For a final summary, our goal was to develop a framework for compositional specification in rewriting logic and Maude and, in the present paper, to show the way for compositional reasoning on such specifications. This much we are confident to have achieved.

We like to think that, through compositionality, rewriting logic can become easier or more suitable to apply to some domains, like runtime verification, coordination models, component-based software development, and hardware specification. All of it is quite speculative at present, which means we have some appealing lines of work ahead of us.

\vspace{6mm}
\noindent\emph{Acknowledgments}: The authors want to thank David de Frutos for his important remark on the correct way to define compatibility of paths, and the anonymous referees for improving this paper with their hard and very useful work.

\vspace{6mm}
\noindent\emph{Competing interests}: The authors declare none.

\bibliographystyle{acmtrans}
\bibliography{compverif}

\end{document}